%% file: elsarticle-template-num.tex
\documentclass[12pt,colorinlistoftodos, table]{elsarticle}

\usepackage{amssymb}
\usepackage{amsmath}
\usepackage{svg}

\usepackage{multirow}
\usepackage{makecell}

\usepackage{rotating}
\usepackage{pdflscape}

\usepackage[colorlinks=true, allcolors=blue]{hyperref}
\usepackage[textsize = small]{todonotes}
\usepackage{glossaries}

\makeglossaries
\input{myacronyms}

\begin{document}

\begin{frontmatter}



\title{Safety Assessment for Autonomous Systems' Perception Capabilities}


\author[inst1]{J F Molloy \corref{cor}}
\ead{john.molloy@york.ac.uk}

\affiliation[inst1]{organization={Department of Computer Science},%Department and Organization
            addressline={Deramore Lane, University of York, Heslington}, 
            city={York},
            postcode={YO10 5GH}, 
            state={Yorkshire},
            country={UK}}

\author[inst1]{J A McDermid}

\cortext[cor]{{Corresponding Author}}

\begin{abstract}
\acrfull{as} are increasingly proposed, or used, in \acrfull{sc} applications. Many such systems make use of sophisticated sensor suites and processing to provide scene understanding which informs the \acrshort{as}['] decision-making. The sensor processing typically makes use of \acrfull{ml} and has to work in challenging environments, further the \acrshort{ml}-algorithms have known limitations,e.g., the possibility of false-negatives or false-positives in object classification.
The well-established safety-analysis methods developed for conventional \acrshort{sc} systems are not well-matched to \acrshort{as}, \acrshort{ml}, or the sensing systems used by \acrshort{as}. This paper proposes an adaptation of well-established safety-analysis methods to address the specifics of perception-systems for \acrshort{as}, including addressing environmental effects and the potential failure-modes of \acrshort{ml}, and provides a rationale for choosing particular sets of guidewords, or prompts, for safety-analysis. It goes on to show how the results of the analysis can be used to inform the design and verification of the \acrshort{as} and illustrates the new method by presenting a partial analysis of a road vehicle. Illustrations
in the paper are primarily based on optical sensing, however the paper discusses the applicability of the method to other sensing modalities and its role in a wider safety process addressing the overall capabilities of \acrshort{as}.
\end{abstract}

%%  Graphical abstract
%%  \begin{graphicalabstract}
%%  \includegraphics{grabs}
%%  \end{graphicalabstract}

%%  Research highlights
%%  \begin{highlights}
%%  \item Research highlight 1
%%  \item Research highlight 2
%%  \end{highlights}

\begin{keyword}
Autonomous Systems \sep Perception Systems \sep HAZOP \sep Machine Learning \sep ALKS
%%\PACS 0000 \sep 1111
%%\MSC 0000 \sep 1111
\end{keyword}

\end{frontmatter}

%%  \section{To Do List}\label{sec:todolist}
%%  To be removed later
%%  \listoftodos 
%%  \linenumbers

\section{Introduction}\label{sec:intro}

\acrfull{as} depend on complex perception subsystems to produce and continuously update a ``world model'' on which to base decisions, e.g. to choose a trajectory. These subsystems are typically multi-modal, i.e. use multiple sensor types, and often employ \acrfull{ml} for processing data from optical sensors, especially cameras. 

The ``world models'' produced by the perception systems are crucial to overall system safety. The decision-making element of the system has to treat the model as ``ground truth'' -- it has no alternative source against which to check the model -- so a ``correct'' decision-making algorithm may give rise to unsafe behaviour if the ``world model'' is sufficiently misleading, e.g. fails to identify an object in the path of a mobile system, such as a road vehicle. Of course, decision-making algorithms themselves can be inappropriate, but our focus here is on unsafe behaviour that can be induced by inadequacies or limitations in the perception subsystem of \acrfull{av}, especially where behaviour relies, at least in part, on \acrshort{ml}.

The paper presents an approach to hazard and safety analysis of \acrshort{ml}-based perception subsystems, as part of a wider programme to adapt classical safety engineering methods for use on \acrshort{as} operating in complex environments. Specifically, it proposes an adaptation of well-established safety analysis methods to address the specifics of perception systems for \acrshort{as}, including addressing environmental effects and the potential failure modes of both sensors and \acrshort{ml}. It focuses on a single domain, \acrshort{av}'s and on camera-based sensing because of their importance and the availability of data, including on critical malfunctions. However, we start by explaining the level of dependency on sensing and present a more extensive rationale for choosing to develop our method based on analysis of camera-based systems. 

\subsection{Importance of Sensing Systems} \label{ssec:importsensingsystems}

Sophisticated sensor systems are now an integral part of consumer vehicles, and 92\% of new cars sold in the US include some \acrfull{adas} features, which are defined as partial automation (Level 2 autonomy) under \acrfull{sae} J3016 \cite{J3016201806Taxonomy, cummingsAutonomousDrivingAssist2021}. \acrshort{adas} features are now widespread, and the first Level 3 features have recently been approved for use in commercial vehicles, e.g. both Honda and Mercedes have received regulatory approval for \acrfull{alks} \cite{ALKS2021}. Typically capabilities at the higher levels of autonomy are referred to as \acrfull{ads}. 

\begin{figure}[h]
    \centering
    \includegraphics[scale=.6]{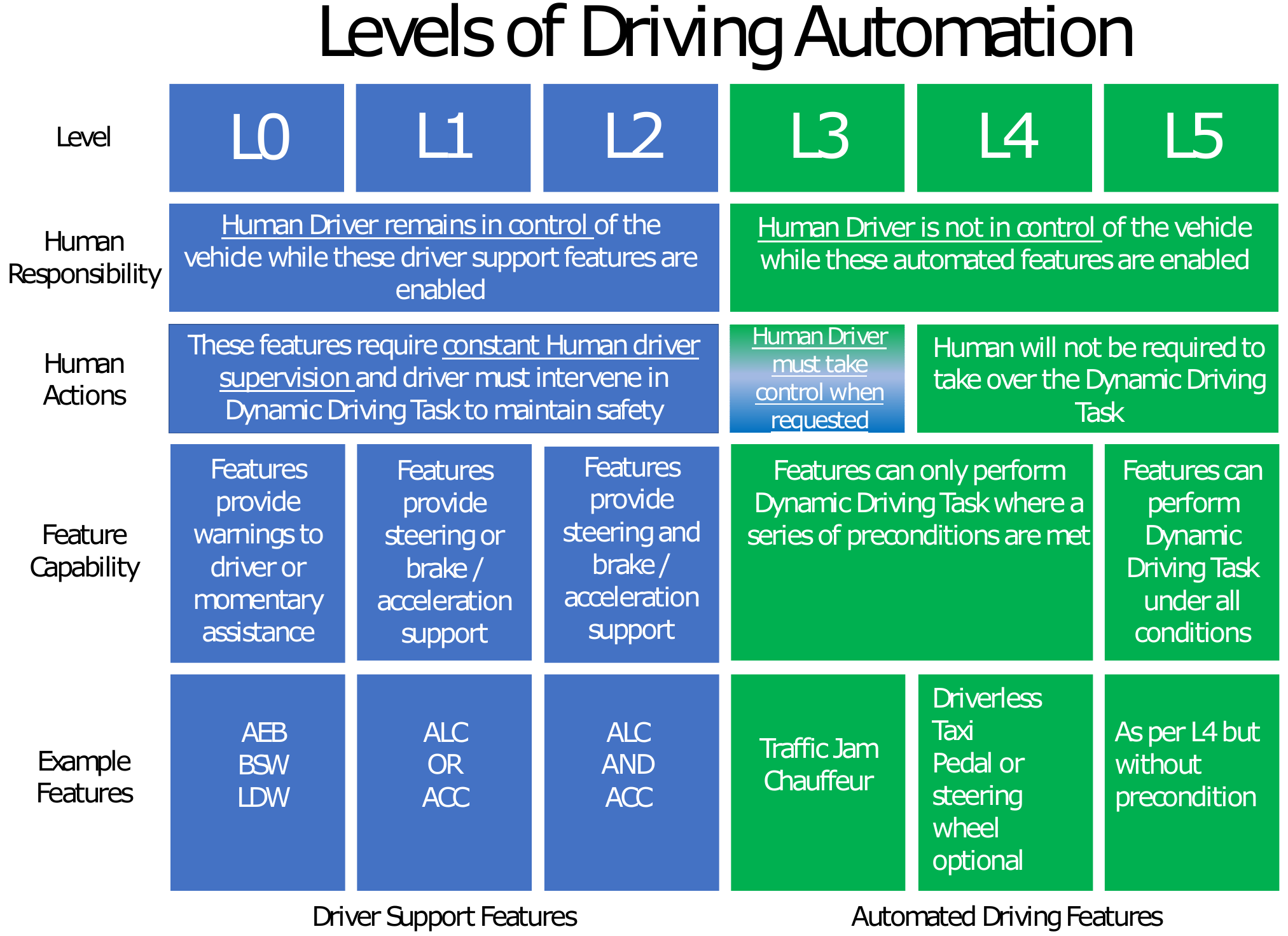}
    \caption{Overview of SAE J3016 adapted from\cite{J3016201806Taxonomy}}
    \label{fig:OverviewOfSAEJ3026}
\end{figure}

Levels of autonomy are defined by the \acrshort{sae} in J3016 depending on the degree of assistance provided and the extent of driver/operator oversight required, see Figure \ref{fig:OverviewOfSAEJ3026}. 
Levels 0-2 require constant driver supervision and the driver remains in control of the vehicle, while Levels 3 and above permit the drivers to engage in activities other than the \acrfull{ddt}. 

The perception system is clearly \acrfull{sc} at Level 3 as, for example, failure to correctly identify lane markings could lead an \acrshort{alks}-equipped vehicle to stray from the current lane. 

\begin{figure}[!h]
    \centering
    \includegraphics[width=1\linewidth]{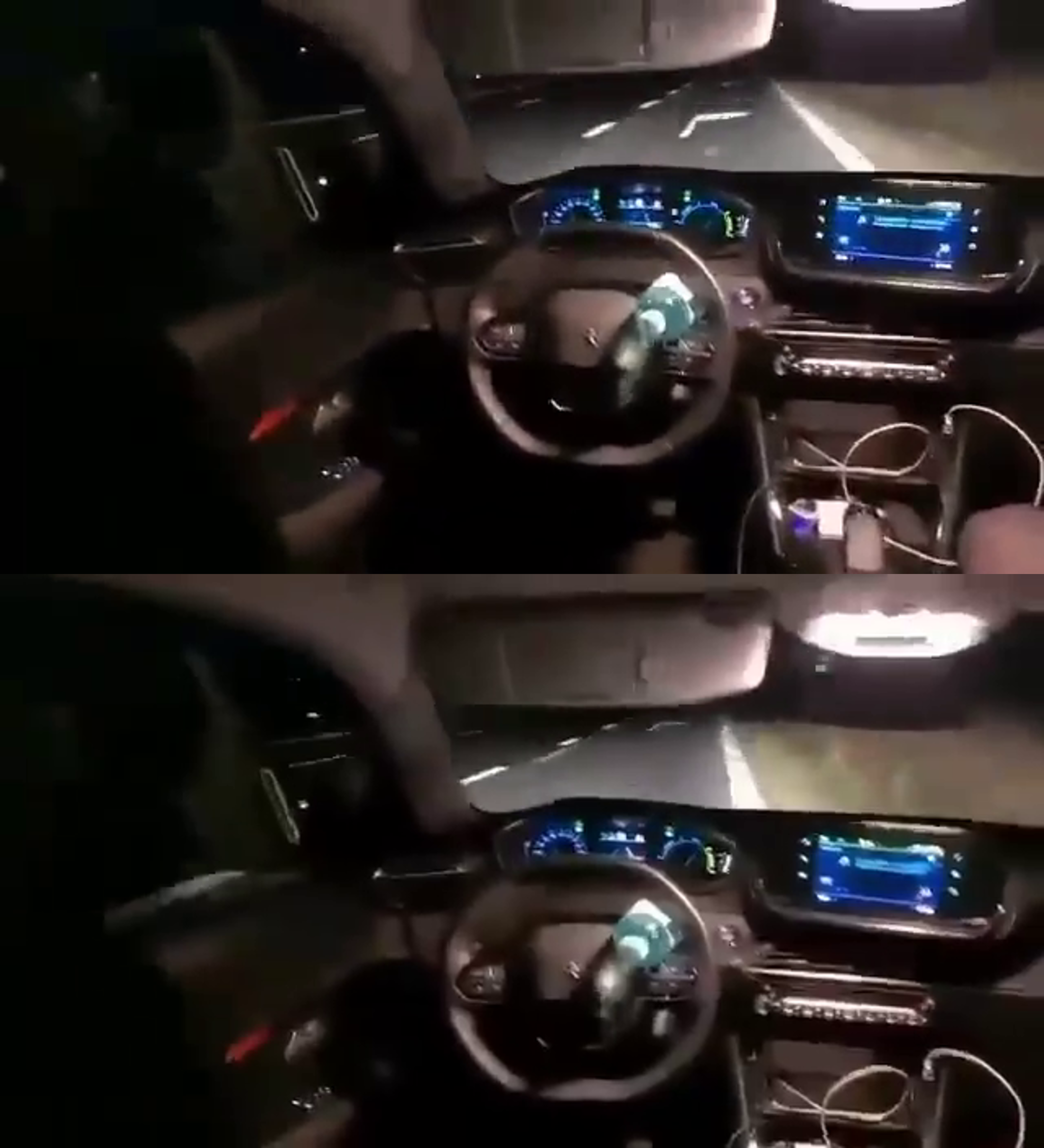}
    \caption{Peugeot 508 on ADAS leaving the road \cite{alexraicu90youtubeSittingBackBragging2021}}
    \label{fig:P508ADAS}
\end{figure}

A recent example shows a Peugeot 508 running an \acrshort{adas} (see Figure \ref{fig:P508ADAS}) leaving the motorway and into the verge, while the driver in the rear seat boasts that a water bottle (used to simulate manually applied steering torque) is driving the car \cite{alexraicu90youtubeSittingBackBragging2021}.

Even for Level 2 systems, errors in the ``world model'' can lead to unsafe behaviour. For example, if an \acrfull{aeb} system initiates braking when there is no obstacle in its path (perhaps due to a false positive from the \acrshort{ml} image processing software) this could give rise to a tail-end collision. With Level 2 systems, drivers are meant to be attentive and able to deal with such situations but, at a minimum, such a malfunction would increase risk. In our \acrshort{hazop} method, we do not use the \acrshort{sae} Level as a factor when considering potential malfunctions of the sensing system, but we illustrate the method using \acrshort{alks}, see Section \ref{sec: ProposedMethod}, where the criticality of misperception is unequivocal. 

Manufacturers have generally taken the approach that vehicles should be self-sufficient in implementing features such as \acrshort{alks}, rather than being dependent on the infrastructure, to for example, communicate the state of traffic lights. Vehicles do this using on-board exteroceptive sensors systems including cameras, radar \& lidar to produce ``world models", in a given \acrfull{odd}. These sensors are used in conjunction with \acrshort{ml} algorithms to detect, understand, and interpret the external environment and its context including: the drivable area, other road users, static obstacles, road signals, e.g. speed limits, traffic lights, etc.

It is difficult, if not impossible, to use a single sensor or data source to accurately interpret complex scenarios such as those encountered by \acrshort{av}'s and to report on all features of interest. Hence, manufactures typically use complementary sets of sensors, fusing their outputs in an attempt to mitigate individual sensor weaknesses \cite{fayyadDeepLearningSensor2020}. 

So, while the ideal perception suite remains an open area of consideration \cite{karpathyCVPR21Workshop2021}, significant benefits are derived from the use of a diverse sensor suite, operating at heterogeneous electromagnetic wavelengths and fusing their outputs: a typical sensor suite now contains multiple \acrfull{rgb} cameras, lidar and radar sets, and additional systems such as thermal imaging, and ultrasound may also be included. Despite the benefits of overcoming the limitations of individual sensing modalities, there are drawbacks in terms of financial cost and computational resources.
\subsection{Hazard and Safety Analysis}
\label{HazardSafetyAnal}

Classical safety processes start by identifying \textit{hazards}, i.e., undesirable situations that pose risk to life, then determine potential \textit{causes} of the hazards and estimate the \textit{risk} associated with the hazards. Typically, risk is a combination of the likelihood of the hazard occurring and the severity of the harm arising from the hazard, although the detailed computations vary from domain to domain, e.g. in the automotive domain \textit{controllability} of the hazard is normally considered.

In the context of this paper, it is assumed that vehicle level hazard analysis has been carried out, incorporating process such as \acrfull{sace} and that the concerns in analysing sensing systems are in identifying potential causes of known hazards we will refer to this as \acrfull{saus} \cite{hawkinsGuidanceSafetyAssurance2022a}. As indicated above, hazards can be caused by errors in the ``world model'' produced by the sensing system. We introduce the concept of a \acrfull{hiss} which arises when the system’s model of the environment or its own state differs significantly from the real world (ground truth). The term ``significantly'' is used as all models will lag the real-world due to the time taken for signals to reach the sensor (or round-trip delays for active sensors) and the length of time for the software to process the sensor data. The models will also necessarily be approximate as sensors have finite resolution, and their precision is also a function of target range. 

The system model must also consider its own state -- for example, whether or not brake actuators are working correctly, as this can also affect safety. Our focus in this paper is mainly on exteroception, but a brief discussion is given of interoception, especially the ability to monitor the state of the sensing system itself.

Identifying hazards and potential hazard causes is a creative process, and many hazard and safety analysis methods use guidewords to prompt the analysts to consider ways in which deviations from design intent might arise and contribute to hazards. To our knowledge, there are no established hazard and safety analysis methods focused on completed ML-driven perception capabilities. Though camera performance for computer-vision has been previously considered \cite{zendelCVHAZOPIntroducingTest2015} showing the methods flexibility, and we believe it is both practical and valuable to adapt existing methods for this purpose.

\subsection{Scope and Structure of the Paper} 
\label{ssec: structure}

For simplicity, and because of the importance of optical perception for human drivers, the paper focuses on camera-based elements of sensing systems from the point of view of identifying potential causes of hazards. Further, we use the term \acrshort{av}s throughout, although we recognise that there are currently no commercially deployed systems at \acrshort{sae} Levels 4 \& 5, and Level 3 systems are just entering the market (although there are highly autonomous systems in other domains). We believe this is reasonable, as it is our intent that the method we propose would be valid and useful, for all of the \acrshort{sae} Levels.\\

The rest of the paper is structured as follows. Section \ref{Sec: background} presents background on the capabilities and limitations of sensing systems with a focus on cameras and the \acrshort{ml} algorithms used for processing the images. It also briefly summarises the principles of classical safety analysis and their role in achieving and assuring system safety. Section \ref{sec: ProposedMethod} sets out our proposed analysis method, which is an adaptation of classical \acrfull{hazop} \cite{kletz2018hazop} and presents examples to illustrate the guidewords chosen for the method. Section \ref{sec: ALKSExample} illustrates the use of the the adapted \acrshort{hazop} method on an \acrshort{alks} capability, making reasonable assumptions about the implementation; treating both camera and radar-based capabilities, to show the generality of the \acrshort{hazop} method. This is followed by a discussion of limitations of the approach and the broadening of the method to deal with wider multi-modal sensing systems, and the role of the method in an overall safety process for \acrshort{as} in Section \ref{sec: Discussion}. Conclusions are presented in Section \ref{sec: Conclusions}.

\section{Background}
\label{Sec: background}

This section first considers the capabilities and limitations of camera-based perception systems, including the \acrshort{ml} algorithms for processing the images, in the context of \acrshort{av}s. This is followed by an introduction to hazard and safety analysis processes, and an assessment of their capabilities and limitations for assessing \acrshort{as}, in particular for the safety analysis of sensing systems. 

\subsection{Capabilities and Limitations of Camera-based Perception Systems} \label{ssec: CameraSystems}

\acrshort{rgb} cameras are to a large extent the principal sensing channel in \acrshort{as} and have become ubiquitous in \acrshort{av}s. Cameras provide reasonably precise and feature rich data, which is important in the context of a road environment designed to be navigated largely by human vision, while also being relatively cheap and robust sensors.

Cameras systems however suffer a number of limitations. While depth data may be extracted from both stereo and monocular images it is generally computationally expensive, of limited accuracy, and best suited to shorter ranges. Additionally \acrshort{rgb} cameras typically perform poorly in low light conditions and, in contrast to human vision, struggle with \acrfull{hdr} scenarios. These weaknesses are typically addressed through multiple exposures and onboard image processing i.e. \acrshort{hdr} imaging, and through the use of complementary sensing modalities i.e. lidar and radar. 

Moreover, visual sensor technology is at an extremely mature stage and total failure of the sensor is now rather rare with significant mitigations in place during production to avoid this. In contrast intermittent failure or processing errors are not as well controlled, and analogue side failures are difficult to identify using automatic software processing at run-time \cite{onsemiconductorEvaluatingFunctionalSafety2018}.

Until recently, the outputs of such sensors have been interpreted by operators to enhance human situational awareness e.g. by increasing visible range or detecting low observable objects. Operators (e.g. vehicle drivers) can often compensate for such errors, intermittent failures, and degradation of the sensor suite by means of training, experience, ability to adjust the sensor set or independent assessment of the scenario. Though interpretation failures still occur, infamously the Pitot tubes on Flight A447 \cite{bea2012accident} and the radar set on USS Fitzgerald (DDG-62) \cite{CollisionUSNavy,CNOUSSFitzgerald2017} it is important to note, that up to \acrshort{sae} Level 2 sensing systems are an aid to the drivers who remain in control of the vehicle and must supervisor it's behaviour, while receiving far less training than the operators in the above two cases.
%\begin{center}\rule{0.5\linewidth}{1pt}\end{center}
%\textbf{JFM Comment:probably extraneous} Attempts at control and automation of vehicles go back to 20s \& 50s, while more recent developments began in the late 1980s onward in Germany and elsewhere seeing projects such as Prometheus, where a Daimler-Benz VaMP, a classically controlled semi autonomous vehicle, which drove journeys in excess of 1000 km
%\begin{center}\rule{0.5\linewidth}{1pt}\end{center}

The advent of \acrfull {cnn} on cheap, powerful computers, e.g. \acrfull{gpu}, has enabled modern implementations of \acrshort{av}s. \acrshort{cnn}'s can provide a ``decent level of accuracy'', and fast processing speeds for many \acrshort{av} tasks including perception in comparison with traditional methods\cite{fayyadDeepLearningSensor2020}. 

Hence has been a sustained increase in use of \acrshort{cnn}s for image processing tasks in \acrfull{as} in the last decade. For example, traffic sign recognition with multi-scale \acrshort{cnn}s is one of the most mature and widely adopted techniques \cite{sermanetTrafficSignRecognition2011}. Whilst it is difficult to get concrete data from manufacturers, research papers produced by manufacturers and their Tier 1 suppliers point to the growing use of such technology, e.g. for pedestrian detection. 

Some \acrshort{av}s also use \acrshort{ml} for other functions, e.g. mapping and decision making, but these applications are outside the scope of this paper. 

%These perception tasks would have previously required significant hand crafting of algorithms and would likely have required much more frequent human interventions in deployment

However, the deployment of \acrshort{cnn}s presents serious safety challenges as there is a significant lack of human interpretability or intuition for the functioning of these \acrfull{nn} \cite{stoccoMisbehaviourPredictionAutonomous2020}, even in the case of visual imagery. This is compounded by the variance in perception suite performance, where it simultaneously out- and under- preforms human capability leading to a cognition gap \cite{huangMixedAnalysisInfluencing2020, cummingsSafetyImplicationsVariability2021} making it difficult for drivers/operators to predict the vehicle's (mis-) behaviour with respect to unseen or unforeseen scenarios and edge-cases \cite{alvarezmelisRobustInterpretabilitySelfExplaining2018}.
Where \acrshort{cnn}s are used for object detection and classification, e.g. distinguishing pedestrians from street furniture, they are prone to \textit{false positives}, e.g. classifying and object as a pedestrian which isn't, and \textit{false negatives}. A false negative which means that an area of a scene is interpreted as a drivable area where, in fact, there is an obstacle (whether a human or static) is clearly hazardous -- a \acrshort{hiss}. More subtly, object classification can be used to shape the \acrshort{av}s prediction of object trajectories, so misclassification can lead to incorrect prediction of the future position of an object -- this was one of the factors in the \acrfull{ntsb}'s investigation of the Uber Tempe fatality \cite{NTSB2016Uber}.

This suggests that the average or even trained drivers may not know whether or why the the system component, either \acrshort{cnn} or sensor, is failing or degraded. Moreover with \acrfull{ota}  updates to core system functionality, the behaviour of the vehicle (or factors affecting the vehicle) may change overnight. Thus human oversight is hampered by significant configuration variability, which was a contributing factor to the fatal USS McCain (DDG-56) (2017) incident \cite{CollisionUSNavya, CNOUSSFitzgerald2017}

Hence, it cannot be reasonable nor ethical to expect the a human driver to provide supervisory behaviour for \acrshort{alks} -- and it could be said to be in conflict with the very definition of the higher \acrshort{sae} levels; in the US \acrfull{nhtsa} preliminary statement of policy concerning automated vehicles (in 2013) the driver cannot be relied upon to provide control methods during operationa of a fully automated vehicle \cite{koopmanChallengesAutonomousVehicle2016}.

In the UK, work by the Law Commission is seeking to identify regulatory changes that would give clarity relating to driver and manufacturer responsibility for \acrshort{av}s at Level 3 and above, with the onus on vehicle manufacturers to ensure safety. In turn, this means that there is a need for hazard and safety analysis of perception systems -- but before we consider that, we first outline some of the uses of \acrshort{ml} in perception systems for \acrshort{av} and the problems that can arise. 

%However, owing to the lack of interpretabiliy/intuitiion it is infeasible to rely on the human driver to provide oversight of sensor and ML performance even at autonomy levels II and III. Though some mitigation could be achieved though a requirement for regular training and evaluation is appears unlikely for large scale deployment with out significant cultural, regulatory and legislative changes.

%This challenge goes beyond the well-known question of human situational awareness, it relates to the inability of trained or otherwise human operator to reasonably understand sensor interaction with external environment and the algorithmic decision based on that data. Therefore the sensors themselves or internal processing architecture must provide supervisory or watchdog type function.

%\begin{center}\rule{0.5\linewidth}{0.5pt}\end{center}

%\begin{center}\rule{0.5\linewidth}{0.5pt}\end{center}

We will use the term \acrfull{ddm} as a general description for the results of \acrshort{ml}, as the issues are not particular to \acrshort{cnn}s \cite{hendrycksNaturalAdversarialExamples2021}. It is difficult (if not impossible) to show deterministically that the outputs of \acrshort{ddm}s will be correct. While \acrshort{ddm}s often produce a measure of confidence, e.g. 93\% certain that an object is a pedestrian, they usually do not provide an explicit consideration of
uncertainty or provide a measure that is not statistically robust.  However, even if the algorithms do not provide specific measures, some degree of uncertainty must be ascribed to the \acrshort{ddm}s in the cases where they are used for safety critical functions \cite{klasFrameworkBuildingUncertainty2020}.

In general, three common sources of uncertainty should be considered for \acrshort{ddm}s: the model fit, the data quality, and the scope compliance \cite{klasFrameworkBuildingUncertainty2020}. Model fit is the inherent uncertainty in the \acrshort{ddm}, Data quality is the \acrshort{ddm}'s uncertainty as a result of it's application to input data obtained in sub-optimal conditions (and greatly affect by sensor performance), and scope compliance is where the model may be applied to the scenarios outside of the intended use \cite{klasUncertaintyWrappersDataDriven2019}.

%(Kläs and Jöckel 2020) In the case of \acrshort{ddm}s or components that is machine learning the 
 
%JFM: It is generated by algorithms extracting implicit patterns within the provided training data sets. What this particular pattern is something of a black box the labels input data can be provided at the actual feature components used to extract of patterns the team to the correct labelling results is unclear.

%\begin{center}\rule{0.5\linewidth}{0.5pt}\end{center}

More broadly, \acrshort{ddm}s often fail to generalise outside their immediate domain of training data distribution and the algorithms are often over-confident on inputs that are rare in the training data set \cite{michaelisBenchmarkingRobustnessObject2019}  Thus the systems are vulnerable to these shifts but humans can't provide oversight -- further they should not be expected to do so for higher \acrshort{sae} Levels.

Finally, functionality provided by \acrshort{ml}, or the expected outcome, is not explicitly specified and implemented by software developers, instead the \acrshort{ddm}s are learnt, and that is a challenge for classical approaches to safety engineering.  

%\textbf{JFM Comment:} check reference {[}Dai and Van Gool, 2018{]} Is there a concept of an adjacent domain?

\subsection{Capabilities and Limitations of Classical Safety Engineering Methods} 
\label{ssec: CapLimHaz_and_Safety}

As mentioned above, classical safety engineering methods and processes are based around the notion of hazards and the risk (likelihood and severity) associated with each hazard. Effective safety processes identify potential causes of hazards early in the system development lifecycle, and identify \acrfull{dsr} to reduce risk. These \acrshort{dsr}s can be intended to reduce the likelihood of the hazard arising, e.g. by using redundant sensing, or by mitigating the consequences, e.g. by using airbags to reduce the effects of collisions. We refer to this as \textit{ensuring} safety. 

Safety engineering also has a role in \textit{assuring} safety, with safety analysis methods used to provide evidence that \acrshort{dsr}s have been met, in some cases. They often have a role in showing that hazard-related risk has been reduced to an acceptable level. In many industries, including automotive, it is common to produce a safety or assurance case to support demonstration that a system is safe to operate, and this can be mandated by standards, e.g. ISO 26262 \cite{RoadVehiclesFunctional2018} in automotive. Our focus in this paper is on \textit{ensuring} safety, and the role of safety engineering methods in understanding potential causes of hazards and thus identifying \acrshort{dsr}s.

Safety engineering processes start at whole system level -- in this case an \acrshort{av} -- and seek to identify potential hazards. Hazards can generally arise from failure to control hazardous substances, e.g. asphyxiating gases, or through inadequate control of energy. In the case of \acrshort{av}s the key issue is control of vehicle kinetic energy -- although we are particularly interested here in how perception system inadequacies can contribute to such hazards. At their simplest, whole system level analysis might ask ``what if'' questions with some simple prompts related to function provision:
\begin{itemize}
    \item \textit{Omission} -- function not provided when intended, e.g. brake not applied when vehicle is stationary (and on a slope so it will start to roll);
    \item \textit{Commission} -- function provided when not intended, e.g.  application of \acrshort{aeb} when there is no obstacle in front of the vehicle;
    \item \textit{Incorrect} -- function provided wrongly, e.g. steering angle insufficient to proceed around a bend and stay on the road. 
\end{itemize}

Of these examples, Commission is most likely to lead to a \acrshort{dsr} which flows down to the perception system, in terms of the (acceptable) likelihood of false positives in detecting objects in the vehicle's path. 

As the design evolves, further analysis will be carried out to identify potential causes of hazards (and perhaps new hazards which become apparent as the system design matures). The chemical process industry developed \acrshort{hazop} \cite{kletz2018hazop} as a method for analysing flows through pipework using guidewords, e.g. \textit{more}, \textit{less}, to flow attributes, e.g. volume, pressure. \acrshort{hazop} has been adopted successfully for computer-based systems, reflecting the fact that it is common to model system architecture in terms of data flows. However, it has been simplified to reflect the absence of attributes of the flow of interest other than the existence/value of data, and the more limited set of potential failure modes. One such example method is \acrfull{shard}, developed in the 1990s \cite{mcdermid1995experience}. 

The focus on failures will continue through the design and development process, ultimately arriving at detailed \acrfull{fmea} on system components, e.g. sensors (see \cite{dhillon1992failure} for a survey of \acrshort{fmea} methods). There are likely to be further \acrshort{dsr}s at this level, e.g. to detect and respond to individual hardware failure modes, but methods such as \acrshort{fmea} also have a role in assurance, e.g. showing that failure behaviour is no worse than assumed in earlier, design-stage, analyses. 

The long-established methods, such as \acrshort{fmea} and \acrshort{hazop}, continue to be applied, but there are more modern methods which build on systems theory, viewing ensuring safety as a control problem, for example Leveson's \acrfull{stpa} \cite{ishimatsuModelingHazardAnalysis2010}. \acrshort{stpa} is becoming increasingly widely used in automotive, but our experience has been that there are limitations in applying it for shared control, such as Level 2 autonomy, but it is possible to enhance \acrshort{stpa} models to deal with these challenges \cite{monkhouse2020enhanced}. 

More fundamentally, the failure-focused methods do not explicitly deal with assessing whether or not the intended functions are safe. There are emerging practices and standards for the \acrfull{sotif} \cite{schnellbach2019development} but it remains unclear whether or not these methods are rich enough to deal with the growing complexity of systems. 

Another fundamental limitation of existing methods and standards is that they do not deal with the specific properties of perception systems, including their failure modes. Associated work has developed a processes for \acrfull{sace} \cite{hawkinsGuidanceSafetyAssurance2022a} and \acrfull{amlas} \cite{hawkins2021guidance}. These are relevant for assuring the safety of the \acrshort{as} as a whole and \acrshort{ml} elements of perception systems, although not the perception sytem or the sensors themselves. Moreover, \acrshort{amlas} starts with identifying the context of use and safety requirements (\acrshort{dsr}s) for the \acrshort{ml} elements of the system. However, we are not aware of any analysis methods that help to identify these \acrshort{dsr}s; this is the ``gap'' that we aim to fill in this paper. 

\section{Proposed Method}
\label{sec: ProposedMethod}

\begin{table}[h!]
\centering
\caption{Classical HAZOP Guidewords 
}
\label{tab: HAZOPGW-table}
\begin{tabular}{|l|l|}
\hline
\textbf{Guideword} & \textbf{Meaning}                       \\ \hline
No or Not          & Complete negation of the design intent \\ \hline
More               & Quantitative increase                  \\ \hline
Less               & Quantitative decrease                  \\ \hline
As well as         & Qualitative modification/increase       \\ \hline
Part of            & Qualitative modification/decrease       \\ \hline
Reverse            & Logical opposite of the design intent  \\ \hline
Other than/Instead & Complete substitution                  \\ \hline
Early              & Relative to clock time                 \\ \hline
Late               & Relative to clock time                 \\ \hline
Before             & Relating to order or sequence          \\ \hline
After              & Relating to order or sequence          \\ \hline
\end{tabular}
\end{table}

Methods such as \acrshort{stpa} have merit for  whole-vehicle level analysis, but with sensing we are dealing with information processing systems and a flow-based model seems more appropriate. In particular, we are interested in identifying potential causes of unsafe internal states -- \acrshort{hiss}, in our terminology. \acrshort{shard} was developed from \acrshort{hazop} for computer-based systems but, implicitly, it assumed that the flows were quite simple, e.g. scalar values. Images from cameras are rich and complex, thus it seems appropriate to go back to the original \acrshort{hazop} definition of guidewords to identify what are appropriate prompts for sensing systems, albeit focusing on information content not considering multiple attributes. We first briefly analyse the standard \acrshort{hazop} guidewords for relevance to the analysis of sensing systems, then give examples to make the intended interpretation more concrete. This is followed by an explanation of how we expect the method to be used. 

\subsection{Analysis of Standard HAZOP Guidewords}

The standard \acrshort{hazop} guidewords are set out in Table \ref{tab: HAZOPGW-table} below. We briefly discuss the guidewords to assess their relevance for sensing system analysis. 

The guidewords are assessed as follows:

\begin{itemize}
    \item \textit{No or Not} -- relevant, this is similar to \textit{Omission} and would include false negatives;
    \item \textit{More} and \textit{Less} -- relevant, this can include false positives, e.g. identifying more objects than exist in the scene through reflections from glass buildings, and false negatives;
    \item \textit{As well as} and \textit{Part of} -- relevant, although may be very similar to \textit{More} and \textit{Less};
    \item \textit{Reverse} -- while not relevant in the meaning of flow from the computers to the sensor we have used it to consider the a sign change in scalar or vector values;
    \item \textit{Other than/Instead} -- relevant, this is similar to \textit{Incorrect} and would include, for example, wrong classification;
    \item \textit{Early} and \textit{Late} -- relevant, although the likelihood of seeing such deviations will depend on the sensor physics;
    \item \textit{Before} and \textit{After} -- there may be mis-ordering of returns from active sensors in multi-path scenarios but, for now, we deem this as not relevant as we are interested in what is in the ``world model'' not the order in which the elements were identified.\\
\end{itemize}

There is some overlap between \textit{No or Not}, \textit{More}, \textit{Less}, \textit{As well as} and \textit{Part of}. It may be, with experience, these could be simplified further or reduced to false positive and false negative. However, \textit{More} and \textit{Less} may remain useful where there are quantitative errors in counting multiple objects of the same class in a given scene. This remains an area for future work, including gaining feedback from experience using the method. 

Experience of malfunctions in perception systems (and their impact on vehicle safety) would also lead us to include \textit{Intermittent} as a guideword, reflecting the case where an object is identified, then missed, then identified again. Whilst this could be viewed as a special case of, say, \textit{No or Not} having the guideword explicitly encourages consideration of such cases. 

\subsection{Proposed Guidewords and Rationale}

The proposed guidewords are set out in Table \ref{tab: newHAZOPGW-table} together with interpretations that are more redolent of the behaviour and failure modes of perception systems, especially those based on cameras with \acrshort{ml} used for image processing and understanding. Illustrative examples of the several of the guidewords are provided below, and an illustration of the use of the method on the \acrshort{alks} is given in Section \ref{sec: ALKSExample}. 

\begin{table}[h]
\centering
\caption{Revised HAZOP Guidewords }
\label{tab: newHAZOPGW-table}
\begin{tabular}{|l|l|}
\hline
\textbf{\textbf{Guideword}} & \textbf{\textbf{Interpretation}}                                                                                                   \\ \hline
No or Not                                    & \begin{tabular}[c]{@{}l@{}}Failure to identify a relevant element of the scene \\ (false negative)\end{tabular}                                     \\ \hline
More                                         & \begin{tabular}[c]{@{}l@{}}Identifying more elements in the scene than are\\  relevant (multiple false positives)\end{tabular}                      \\ \hline
Less                                         & \begin{tabular}[c]{@{}l@{}}Identifying fewer elements in the scene than are \\ relevant (multiple false negatives)\end{tabular}                     \\ \hline
As well as                                   & \begin{tabular}[c]{@{}l@{}}Identifying element in the scene that is not there\\ (false positive)\end{tabular}                                       \\ \hline
Part of                                      & \begin{tabular}[c]{@{}l@{}}Failing to identify element in the scene that is there \\ (false negative)\end{tabular}                                  \\ \hline
Other than/Instead                           & \begin{tabular}[c]{@{}l@{}}Incorrect classification, e.g. static object rather than \\ pedestrian\end{tabular}      
    
    \\ \hline
Reverse                           & \begin{tabular}[c]{@{}l@{}}Change of sign in a scalar or vector value, e.g. pedestrian \\ is moving towards rather than away from ego vehicle\end{tabular}  

    \\ \hline
Early                                        & \begin{tabular}[c]{@{}l@{}}Object identified earlier than necessary for safe \\ behaviour, perhaps triggering unnecessary response\end{tabular}     \\ \hline
Late                                         & \begin{tabular}[c]{@{}l@{}}Object identified later than necessary for safe\\ behaviour\end{tabular}                                                 \\ \hline
Intermittent                                 & \begin{tabular}[c]{@{}l@{}}Element of scene present in some images, but not in\\ others, or classification changes from image to image\end{tabular} \\ \hline
\end{tabular}
\end{table}

\begin{figure}[!h]
    \centering
    \includegraphics[width=1\linewidth]{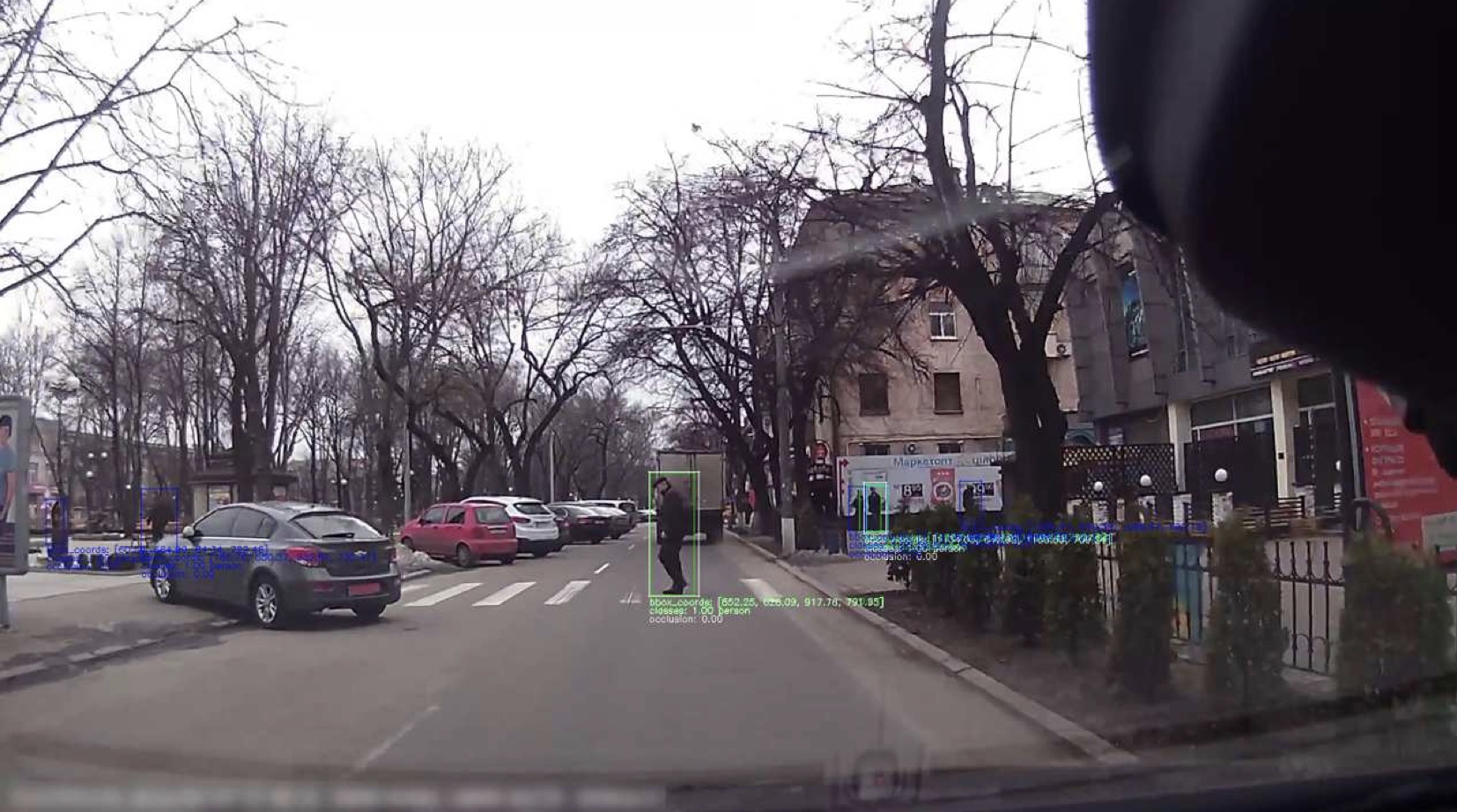}
    \caption{Missing Pedestrians at a Pedestrian Crossing}
    \label{fig:Missing Pedestrians}
\end{figure}

The guideword \textit{No or Not} can be illustrated by an example of pedestrian detection at a level crossing. Figure \ref{fig:Missing Pedestrians} from \cite{gauerhof2020assuring} shows pedestrians correctly detected in green bounding boxes and those that have been missed (false negatives) with blue bounding boxes. One pedestrian is partly occluded by another who is correctly identified; two others are partly obscured by trees. This also serves as an example of \textit{Part of}. Arguably, behaviour in this case is still safe as the pedestrian most exposed to risk has been correctly identified.

\begin{figure}[!h]
    \centering
    \includegraphics[width=1\linewidth]{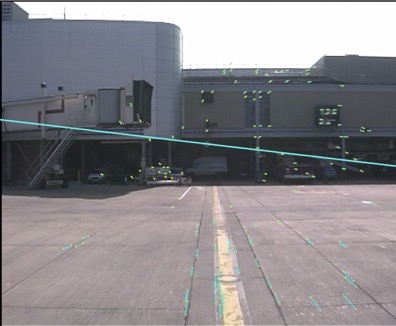}
    \caption{Detection of Spurious Lines in Image for Aircraft Taxiing System }
    \label{fig:SpuriousLInes}
\end{figure}

Figure \ref{fig:SpuriousLInes} is from a system intended to detect runway and taxiway centre lines to assist in landing and aircraft manoeuvres in adverse weather conditions. As well as detecting the painted yellow lines it has ``picked up'' the joins in the concrete, thus illustrating the \textit{More} guideword. 

\begin{figure}[!h]
    \centering
    \includegraphics[width=0.6\linewidth]{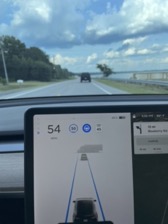}
    \caption{Spurious Object Identification}
    \label{fig:SpuriousObject}
\end{figure}

Figure \ref{fig:SpuriousObject} shows a Tesla system ``seeing'' a sky/cloud formation as a crossing truck (this is apparent on the image of the screen in the vehicle). This is an example of \textit{As Well As}. It can also be seen as an example of pareidolia \cite{abbasFaceRecognitionFacial2019} where \acrshort{nn}s tend to ``see'' patterns in images which are not actually there \cite{ nguyenDeepNeuralNetworks2015, nguyenMultifacetedFeatureVisualization2016, InceptionismGoingDeeper}. In this particular case, the deviation from intent is unlikely to be hazardous (the spurious image will disappear as the vehicle proceeds).

\begin{figure}[!h]
    \centering
    \includegraphics[width=0.5\linewidth]{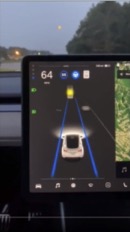}
    \caption{Moon Misclassified as Amber Traffic Light}
    \label{fig: YellowMoon}
\end{figure}

Figure \ref{fig: YellowMoon} shows a Tesla interpreting the moon as an amber traffic light, and this illustrates \textit{Incorrect} -- an inappropriate classification. Depending on the road situation, and the estimated proximity of the vehicle to the identified traffic light the vehicle could start to slow down -- this could be hazardous as other (manually driven) vehicles would not be expecting deceleration for a non-existent traffic light, and there is a risk of a rear-end collision. 

\begin{figure}[!h]
    \centering
    \includegraphics[width=0.8\linewidth]{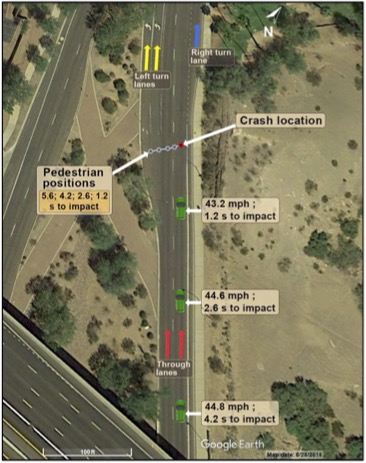}
    \caption{Timeline Leading to Accident with Uber ATG vehicle in Tempe Arizona}
    \label{fig: UberTempe}
\end{figure}

The timing guidewords -- \textit{Early} and \textit{Late} -- are not easy to illustrate using images, instead they would need image sequences or videos. However, the identification of Elaine Herzberg as a pedestrian in the Uber Tempe accident only 1.2$s$ before impact is an example of a \textit{Late} failure (deviations) \cite{NTSB2016Uber}. Further, the Uber systems repeatedly re-classified Ms Herzberg in the seconds leading up to the accident. Key points in the timeline are illustrated in Figure \ref{fig: UberTempe}. This is an example of the \textit{Intermittent} deviation. However, it also shows the need to undertake analysis in the context of understanding system behaviour. At the time of the accident the Uber software discarded the trajectory history for the ``object'' whenever it was reclassified. Thus, for example, the motion was at one time predicted to be along the left turn lanes marked in the figure and not across the road. 

The examples illustrate the guidewords and provide further motivation for their choice. However, it also illustrates that not all deviations from design intent are hazardous, and that knowledge of system design and of operational context are needed to determine whether or not a given deviations constitutes a \acrshort{hiss} and can contribute to a hazard. 

\subsection{Applying the Method}

In broad terms, it is expected that the system architecture can be viewed as four main elements: \textit{Sense}, \textit{Understand}, \textit{Decide} and \textit{Act}. In analysing perception systems we treat \textit{Sense} and \textit{Understand} together, and \acrshort{hiss} can arise at the interface between \textit{Understand} and \textit{Decide}. At this point, the \textit{Decide} element has to ``trust'' the system model and significant deviations between ground truth and the system's model of the environment can lead to unsafe decisions. Several of the above examples make this clear, but it is perhaps most apparent with the Uber Tempe example \cite{NTSB2016Uber}. 

As indicated earlier, the expectation is that hazard analysis will have been undertaken at whole system level (for an \acrshort{av}, in this case), prior to applying \acrshort{saus}. To be of most value, \acrshort{saus} would be applied once a perception system architecture has been defined with commitments made to a particular sensor suite and to the \acrshort{ml} approach, e.g. the structure of \acrshort{ddm}s, etc. so the analysis can accurately  reflect the system design. \acrshort{dsr}s might include refinements to the \acrshort{ddm}'s architecture, adaptations of the hyperparameters, e.g.loss functions, and choice of training data sets. 

Due to the complexity of the \acrshort{adas} or \acrshort{ads} to be analysed, \acrshort{saus} is applied on scenarios. This has the benefit of identifying the \acrshort{ddt} being undertaken by the \acrshort{adas} or \acrshort{ads} and provides sufficient context to assess the potential for hazards. Each scenario is described as a form of use-case including the following elements \cite{guiochetHazardAnalysisHuman2016}:

\begin{itemize}
    \item \textit{Primary environment} -- the context for the \acrshort{ddt}, e.g. type of road;
    \item \textit{Goal in context} -- the top-level goal of the \acrshort{ads} or \acrshort{adas} in this context;
    \item \textit{Scope} -- any restrictions on the scenario, e.g. weather conditions under which the \acrshort{ads} or \acrshort{adas} is expected to operate; 
    \item Pre-conditions -- necessary conditions for the scenario, e.g. presence of other vehicles;
    \item \textit{Success end conditions} -- the results of correct behaviour of the system in the scenario;
    \item \textit{Failed end conditions} -- the hazard(s) identified by earlier analysis;
    \item \textit{Actors} -- the systems and/or individuals involved in the use case, e.g. an \acrshort{adas} and the driver;
    \item \textit{Trigger} -- a condition that can initiate the use case;
    \item \textit{Description} -- the sequence of actions that should occur in the scenario. 
\end{itemize}

In practice, it is likely that the \textit{Primary environment} and the \textit{Scope} will reflect the \acrfull{odd} for the \acrshort{ads} or \acrshort{adas}. 

As with other \acrshort{hazop}-based methods, the analysis is recorded in tables. The key entries in the table are as follows:

\begin{itemize}
    \item Function -- the autonomous capability, i.e. the \acrshort{adas} or \acrshort{ads} function, being analysed;
    \item Parameter -- the information supporting the capability being analysed, e.g. distance to vehicle in front for \acrshort{aeb};
    \item Guideword -- the \acrshort{hazop} guideword to be applied to the parameter;
    \item Deviation -- the interpretation of the guideword for the parameter being considered;
    \item Hazard -- description of the hazard(s) that can arise from the deviation, if any;
    \item Situation -- the context in which the \acrshort{ddt} is taking place, e.g. type of road, reflecting the \textit{Primary environment};
    \item Consequences -- the potential effects of the hazard(s) which may be left blank if the deviations are not considered hazardous;
    \item Causes -- credible malfunctions that could give rise to the identified deviations;
    \item Derived Safety Requirements -- \acrshort{dsr}s where deemed appropriate to reduce the likelihood of the hazard causes or to mitigate the consequences.  
\end{itemize}

The analysis would consider the combination of these elements. There is a risk of ``combinatorial explosion'' but choosing an appropriate level at which to model the system functions, and noting when a guideword produces the same results as identified previously, e.g. considering \textit{Less} after considering \textit{No or Not} may not identify any further deviations, should help to keep the analysis manageable.

\section{ALKS Example}
\label{sec: ALKSExample}

\begin{figure}[ht!]
\label{Fig: ALKS}
  \centering
  \includegraphics[width=\textwidth]{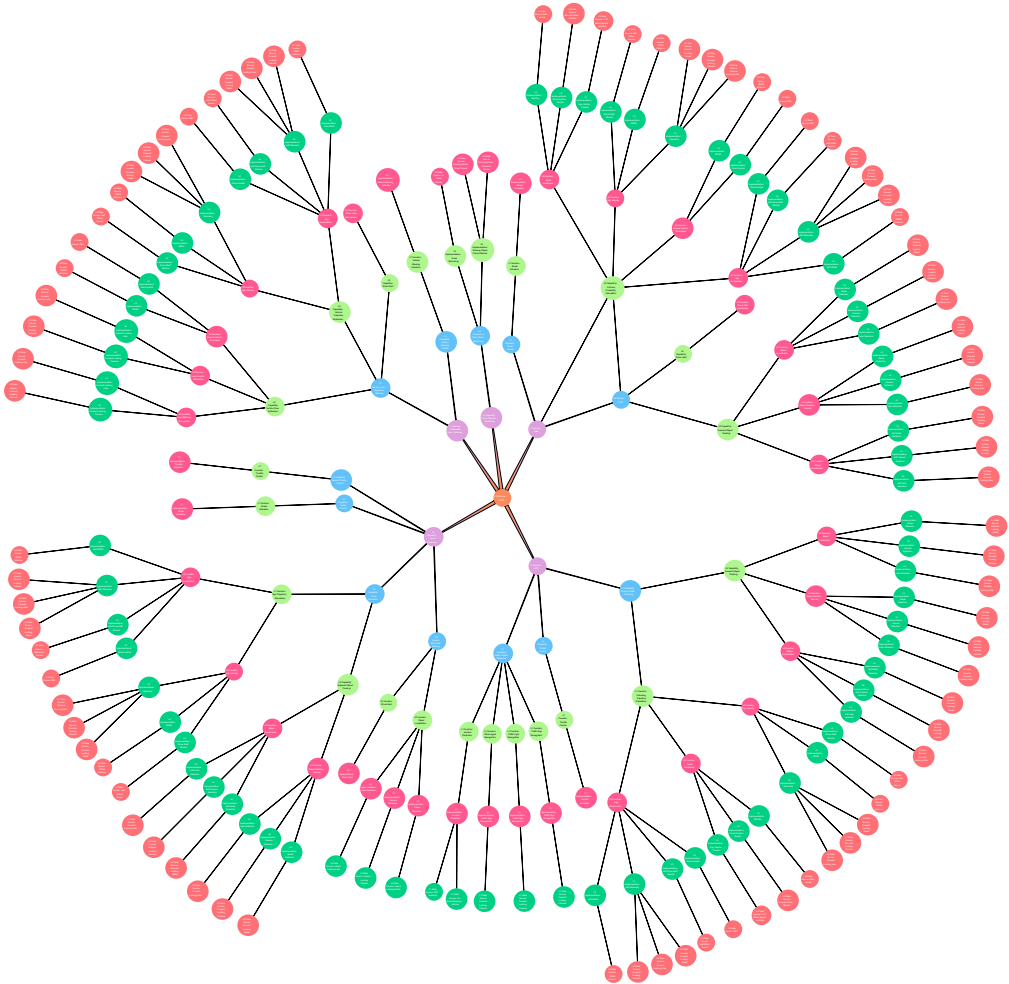}
  \caption{ALKS system, subs-systems and components}
  \label{fig: ALKSmodel}
\end{figure}

Analysis starts with a system model. Here we wish to look at the behaviour (of \acrshort{alks}) at a  high level to elucidate the  the potential hazards and fragility of \acrshort{alks} implementations owing to Sensor vulnerabilities without restriction by the exact sensor device or implementation algorithm. We choose this level as we believe it is useful and informative; see Section \ref{sec: Discussion}
 for a discussion.
 
We decompose the \acrshort{adas} system in this case L2/L3 \acrshort{alks} capabilities into different levels:

\begin{itemize}
    
\item \textbf{Service} -- a complete vehicle feature available to the driver/operator regardless of autonomy level e.g L2+: \acrshort{alks}, L1: \acrfull{acc}, \acrfull{alc} or \acrfull{aeb}, L0: \acrfull{bsw}, \acrfull{fcw} or \acrfull{ldw}

\item \textbf{Behavioural Capability} -- top level capabilities required to implement the service, e.g. ability to change lane

\item \textbf{Function} -- processing of data streams required to implement the capability

\item \textbf{Implementation} --  hardware and software required collect and process data to service the function or give effect to decisions

\item \textbf{Data Source} -- description of unit providing %probable 
data to implementation
 
\end{itemize}

A L2 service may be made up of L1 services in addition to L2 capabilities while a L1 service may be made from L0 services and L1 capabilities. 

Level of service restricts capabilities, e.g. a L1 service permits automated control of steering or acceleration while L2 or above is required for automated control of both steering and acceleration.

This decomposition is illustrated in Figure: \ref{Fig: ALKS} we will concern ourselves with the vertices between Behavioural Capability and Function in order to maintain a high level assessment.\\

\subsection{Illustrations}\label{illustrations}

We will consider several instances taken from real events where a vehicle operating an \acrshort{alks}-like system experiences what we believe to be a \acrfull{hiss}. A \acrshort{hiss} as defined above, is where system’s model of the environment or its own state differs significantly from the real world (ground truth). This difference can result from failures at different levels with in the autonomous system namely in: sensing, perception, understanding and decision-making.

Sensing failures fall into 3 categories: the sensor is non-functional, the sensor is functioning but degraded, or the sensor is malfunctioning. These may arise from a multitude of sources including: mechanical failure, electronic failure, thermal failure, environmental conditions or configuration error, to name but a few.

Whilst it is impossible to understand the environment perfectly with a degraded sensor, more frequently the available data suggests that the failures reside in the interpretation of the provided data leading to perception failure which results in understanding failures. These to fall into five broad categories:
\begin{itemize}
    \item Unrecognised Sensor Failure -- Failure, degradation or malfunctioning of sensor or sensor data is not recognised
    \item Not detected -- Objects appear in the sensor field of view and data but are not detected (Object presence)
    \item Not classified -- Objects are detected but are not classified and their presence is rejected (Object presence)
    \item Misclassification -- Objects are detected but are misclassified (Object nature including pareidolia)
    \item Illusion -- Objects are are detected where they don't exist or their nature is grossly misinterpreted ( Object presence, shape, position, movement, or colour)
\end{itemize}

Perception failures may lead to understanding failures if we consider that understanding is the linking of pieces of information to create a coherent scene. Observed failures of understanding have arisen where there is an apparent conflict between different sources of information or where several pieces of concurring information conflict with the external environment. These fall into several broad categories:

\begin{itemize}
    \item Unrecognised Perception failure
    \item Conflict between sensors' perception 
    \item Conflict between perception and internal law  
    \item Conflict between perception and internal data source
\end{itemize}

We can illustrate these cases cases with some simple examples.

\begin{itemize}
    \item Conflict between sensor data -- detection and classification of targets does not correspond between sensor units.
    \item Conflict with on board law -- radius of curve is perceived as greater than actual, lateral acceleration will exceed maximum permitted limit, $3m/s^{2}$ in normal operation, or $5m/s^{2}$ in an emergency manoeuvre \cite{RegulationNo792018}.
    \item Conflict with on board data -- the detected on the ground geometry of the driveable area, does not match with the driveable area described by internal/on-board maps
\end{itemize}

Finally understanding failures may be driven by decision making, where a higher level decision may override the understanding developed from perception, or to command an action that ignores the developed understanding. This can be illustrated by a commonly observed example:

\begin{itemize}
    \item Conflict between Decision and Understanding -- Decision level route planning overrides understanding to command a manoeuvre that conflicts with the developed understanding such as cutting back to a straight-on route after beginning a right turn.
\end{itemize}

An \acrshort{alks} system principally performs the \acrshort{ddt} by keeping the vehicle in it's lane by control of the lateral and longitudinal movements of the vehicle \cite{UniformProvisionsConcerning2019}. Some currently available examples of \acrshort{alks}-like capability further permit: lane change, junction negotiation and navigate to destination capabilities, without issuing a transition demand.

Examining the requirements for this functionality several of the principal behavioural capabilities are need to create an \acrshort{alks} service namely: \acrshort{acc} \acrshort{alc} \acrshort{aeb} \acrshort{fcw} \acrshort{ldw}. In the main body of the paper we will treat:\acrshort{acc} and \acrshort{alc} via the \acrshort{hazop} methodology further \acrshort{alks} capabilities (\acrshort{aeb} \acrshort{fcw}  \acrshort{ldw}) will be treated in annexes: \ref{sec:appendix1}.

To carry out the \acrshort{hazop} we need an understanding of limits on vehicle dynamics. In this case, it is sufficient to understand limitations on cornering. 
A vehicle navigating a curve experiences a lateral acceleration $a = \frac{v^{2}}{r}$ where $v$ is the vehicles velocity and $r$ is the radius of curvature.
\emph{UN Regulation No. 79: Automatically Commanded Steering Function} specifies that the magnitude of lateral acceleration should not exceed $a > 3 m/s^{2}$ and under nominal consideration the manufacture may permit a maximum of $a > 4m/s^{2}$  \cite{RegulationNo792018}. 

\begin{figure}[ht]
  \centering
  \includegraphics[width=\textwidth]{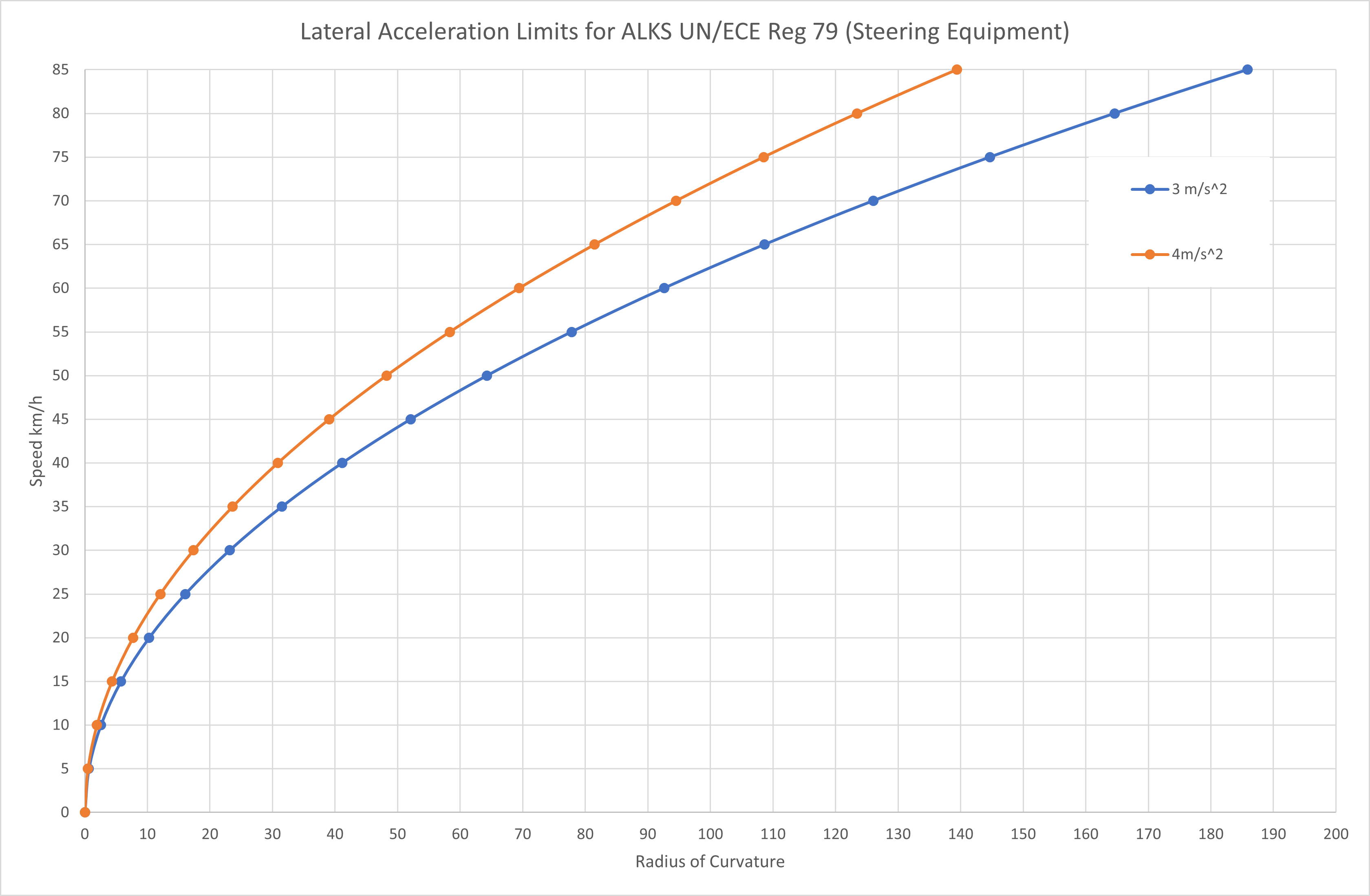}
  \caption{UN Regulation No. 79: Automatically Commanded Steering Function maximum lateral acceleration}
  \label{fig: UneceMax}
\end{figure}

When applying the \acrshort{hazop} Guide Words to parameters of \acrshort{adas} functions Reverse results are considered in some instances e.g. \acrshort{acc} 1.1.2 \emph{Distance to Target} (see Table \ref{fig: Hazop_ACC_1}), the finding of a Reverse result in a forward looking sensor could be excluded or mitigated by simple logic check against nonphysical results, and flag an error state in the sensor or software status.

Example \acrshort{hazop} analyses are provided in Tables \ref{fig: Hazop_ACC_1} to \ref{fig: Hazop_ALC_5} below, noting that the \acrshort{dsr}s are left out and are discussed below. We briefly illustrate some of the entries to further explain the method. As noted above, although the definition of the \acrshort{hazop} guidewords was motivated by analysis of camera-based systems we illustrate the method with both camera and radar-based capabilities to help to show its generality. 

For entry \acrshort{acc} 1.1.1 \emph{Relative Velocity of Target Vehicle}, the first row for ``More'' illustrates how the effect of the deviation can vary in different driving situations. Whilst, in itself, this is not hazardous (the distance to the lead vehicle increases) reaction of other road users might contribute to hazards in the wider road transport system, as shown in the consequences column. Here the \acrshort{dsr} is essentially for correct performance -- within some error. There is value in setting expectations on such performance for safety and to enable Tier 1 manufacturers to produce \acrshort{av} components at a reasonable price and often \acrshort{unece} regulations will provide the detail. 
For entry \acrshort{acc} 1.2 \emph{Distance to Target Vehicle}, the last row for ``Reverse'' identifies a situation that is physically impossible. This should lead to a system-specific \acrshort{dsr} to detect and reject such data. Whilst this may seem obvious, a previously analysed a medical device which caused two fatalities, at least in part because it failed to respond to ``impossible'' values. Again, Tier 1 suppliers may produce components with such properties, but the analysis gives a clear basis for checking this. 

Entry \acrshort{fcw} 0.2.5 \emph{Collision Warning} \ref{fig: Hazop_FCW_3} shows how the lower-level deviations come together to cause problems in the generation of the Collision Warning (the causes are deviations in the parameters in entries 1.1 to 1.4). \acrshort{dsr} would also be expected at this level -- for example, ``Not'' would lead to requirements for sensors to self-diagnose their failures (interoception) or for cross-checks between sensors.  

\begin{table}[ht]
\caption{Scenario use-case: Camera implementation of Automated Lane Centring}
\begin{center}
\scalebox{0.70}{
\input{Tables/T_CMRA_ALC_1_Table}}
\end{center}
\end{table}

\begin{landscape}
\begin{table}[ht]
  \centering
  \includegraphics[scale=0.75]{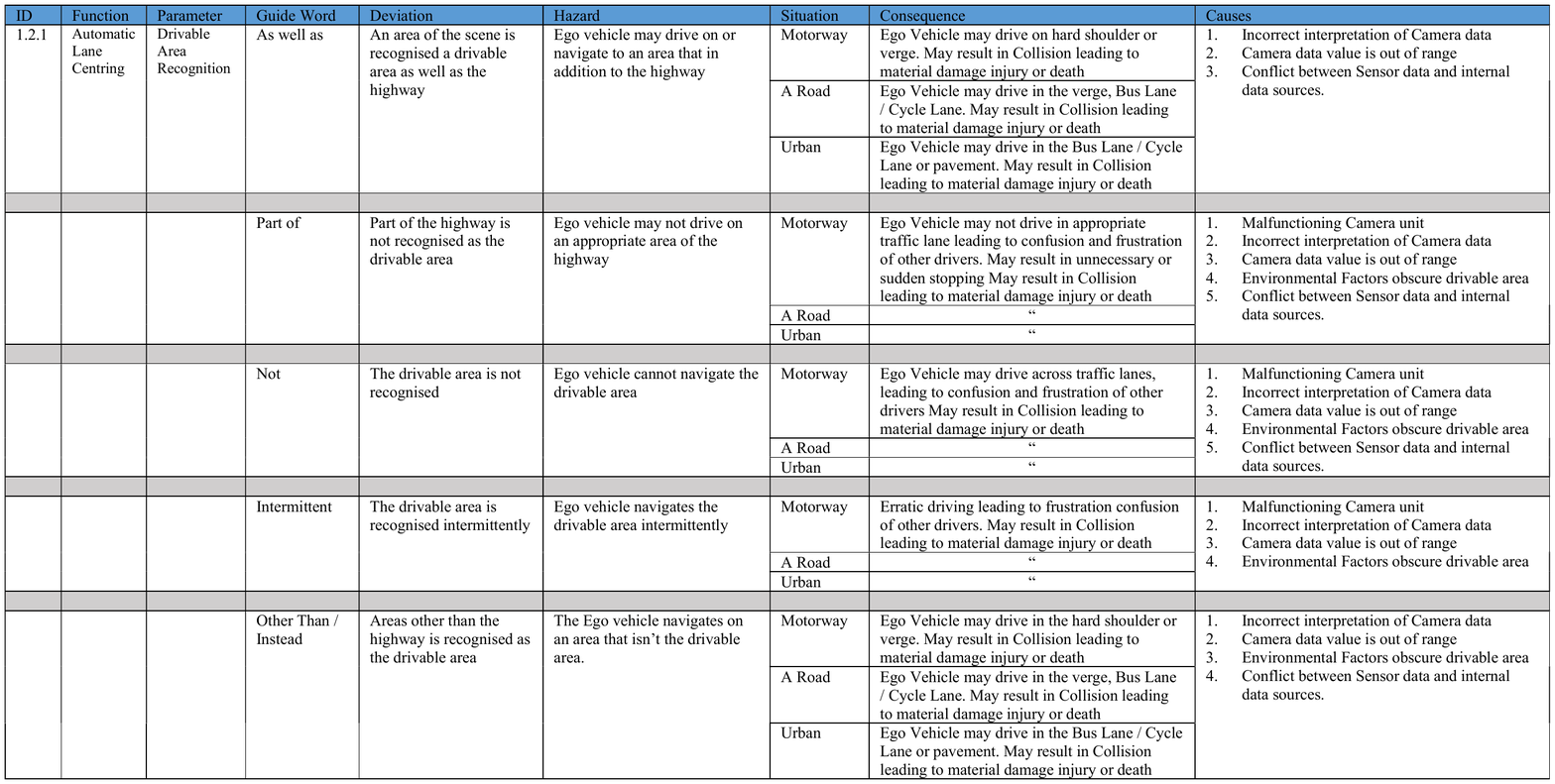}
  \caption{Hazop Table for camera implementation of \acrfull{alc} Part 1}
  \label{fig: Hazop_ALC_1}
\end{table}
\end{landscape}

\begin{landscape}
\begin{table}[ht]
  \centering
  \includegraphics[scale=0.75]{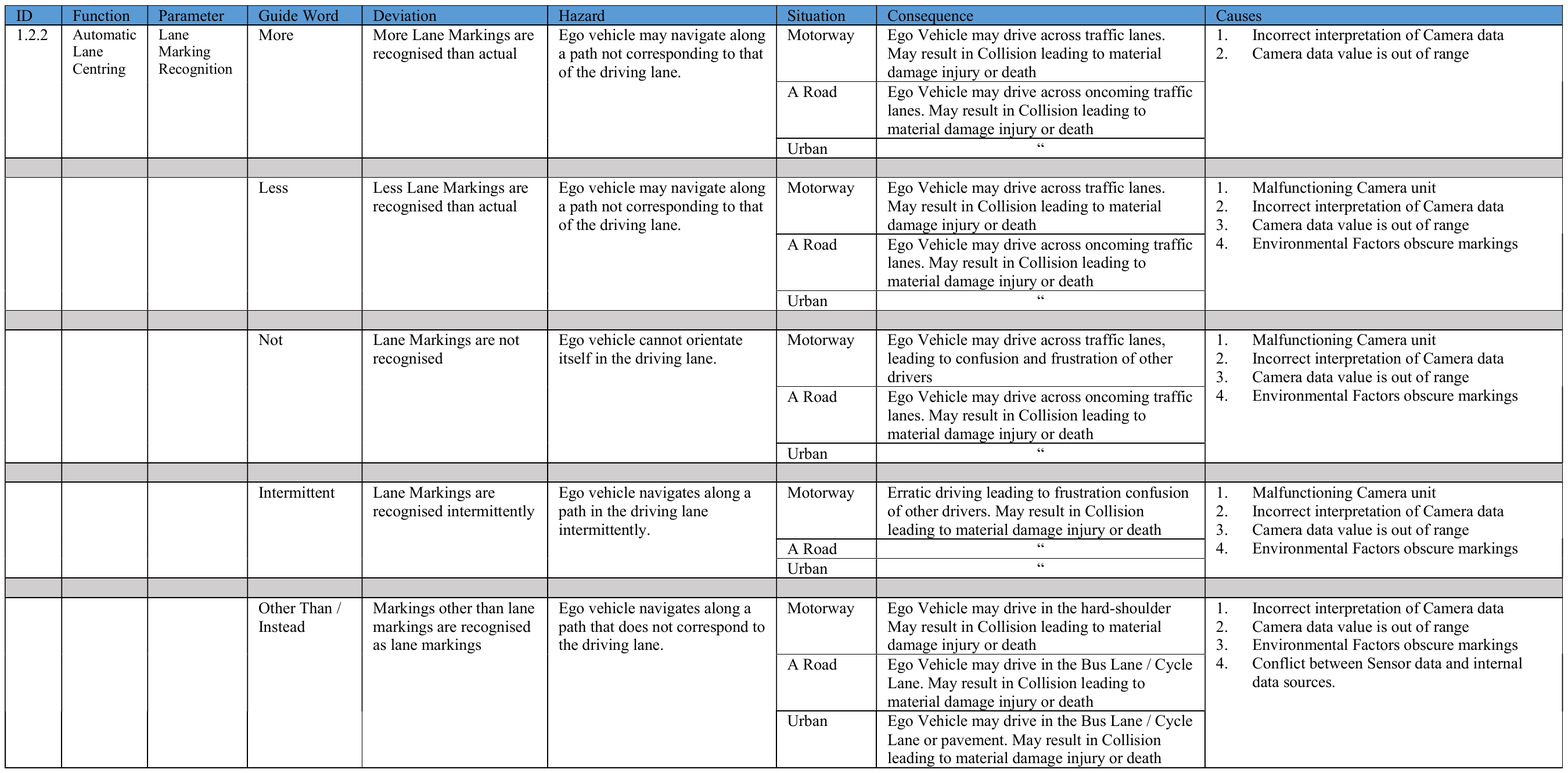}
  \caption{Hazop Table for camera implementation of \acrfull{alc} Part 2}
  \label{fig: Hazop_ALC_2}
\end{table}
\end{landscape}

\begin{landscape}
\begin{table}[ht]
  \centering
  \includegraphics[scale=0.75]{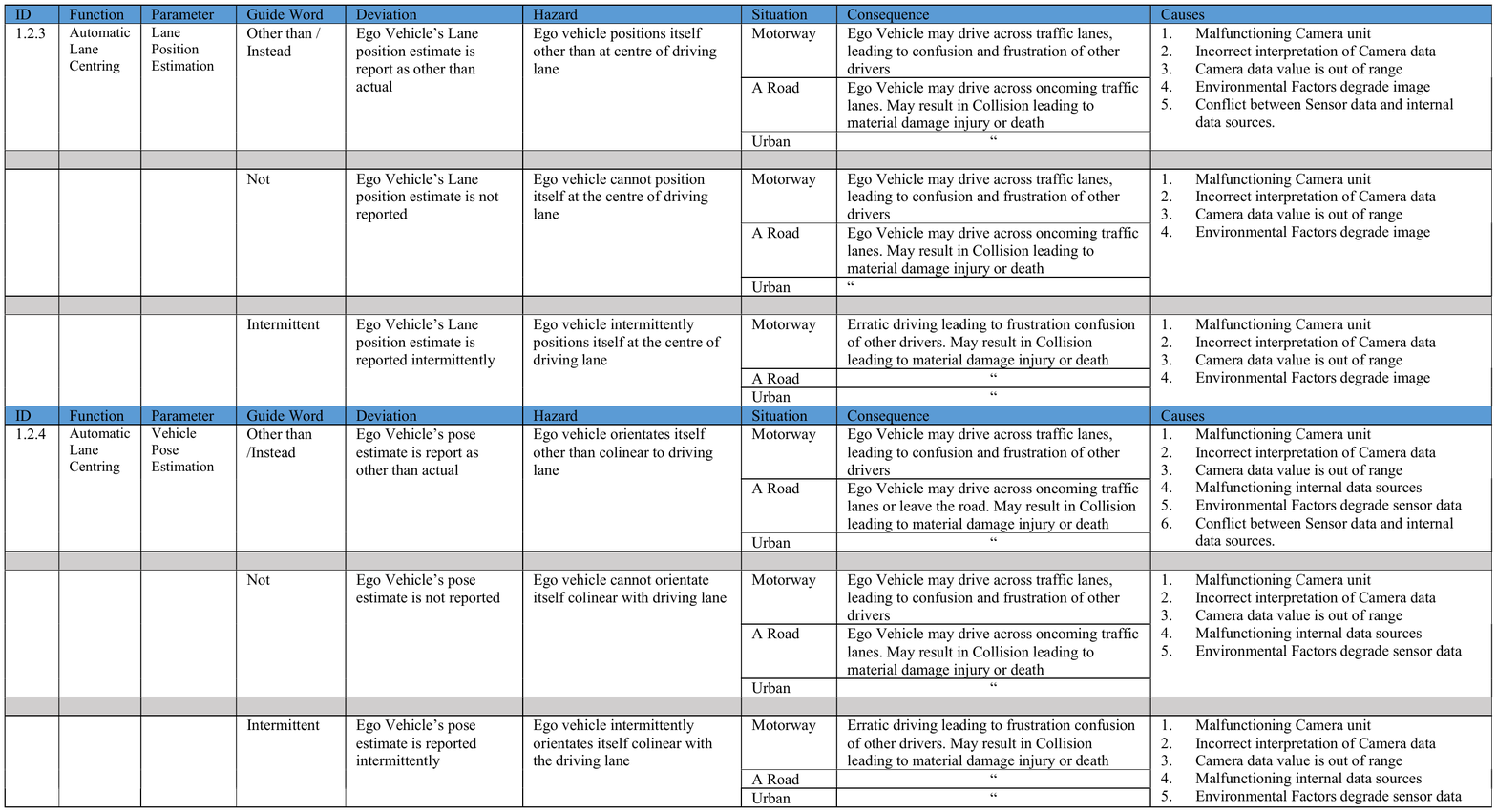}
  \caption{Hazop Table for camera implementation of \acrfull{alc} Part 3}
  \label{fig: Hazop_ALC_3}
\end{table}
\end{landscape}

\begin{landscape}
\begin{table}[ht]
  \centering
  \includegraphics[scale=0.75]{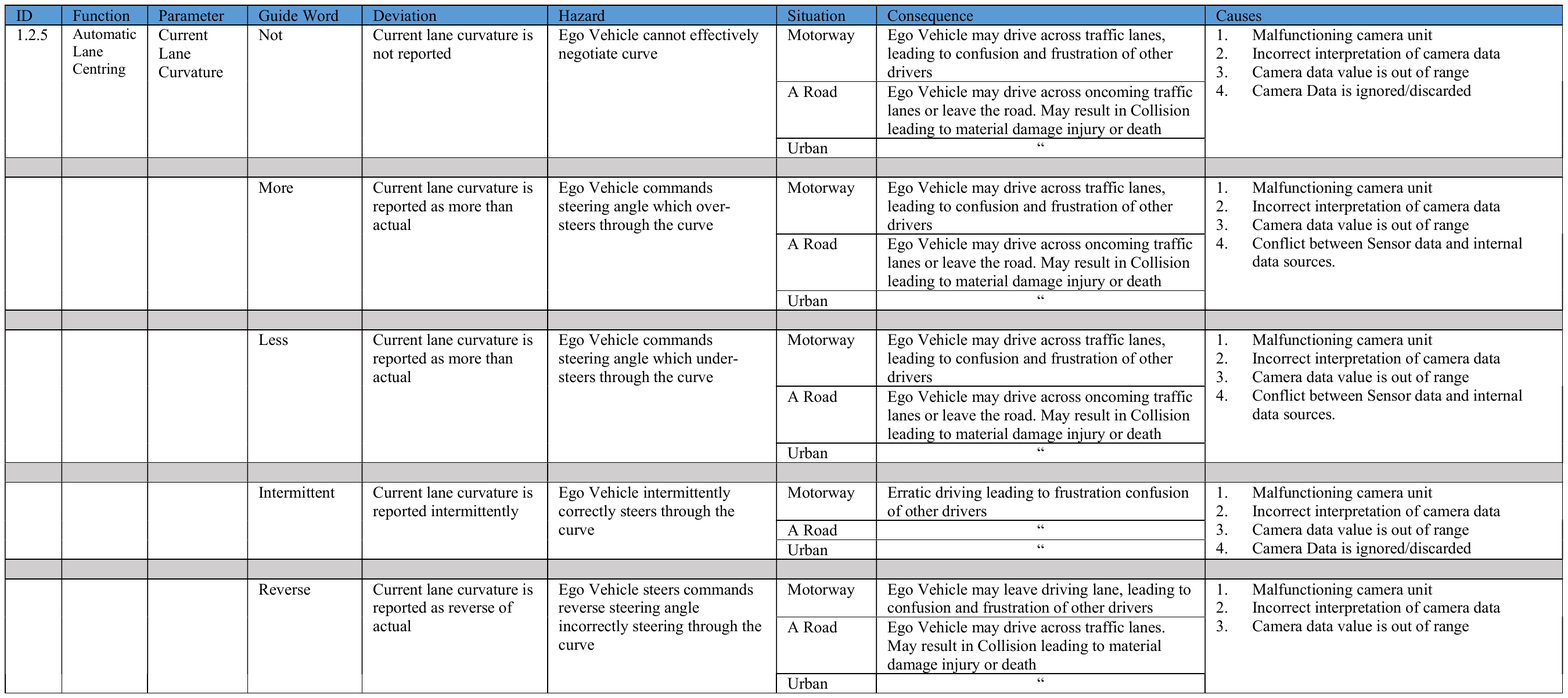}
  \caption{Hazop Table for camera implementation of \acrfull{alc} Part 4}
  \label{fig: Hazop_ALC_4}
\end{table}
\end{landscape}

\begin{landscape}
\begin{table}[ht]
  \centering
  \includegraphics[scale=0.75]{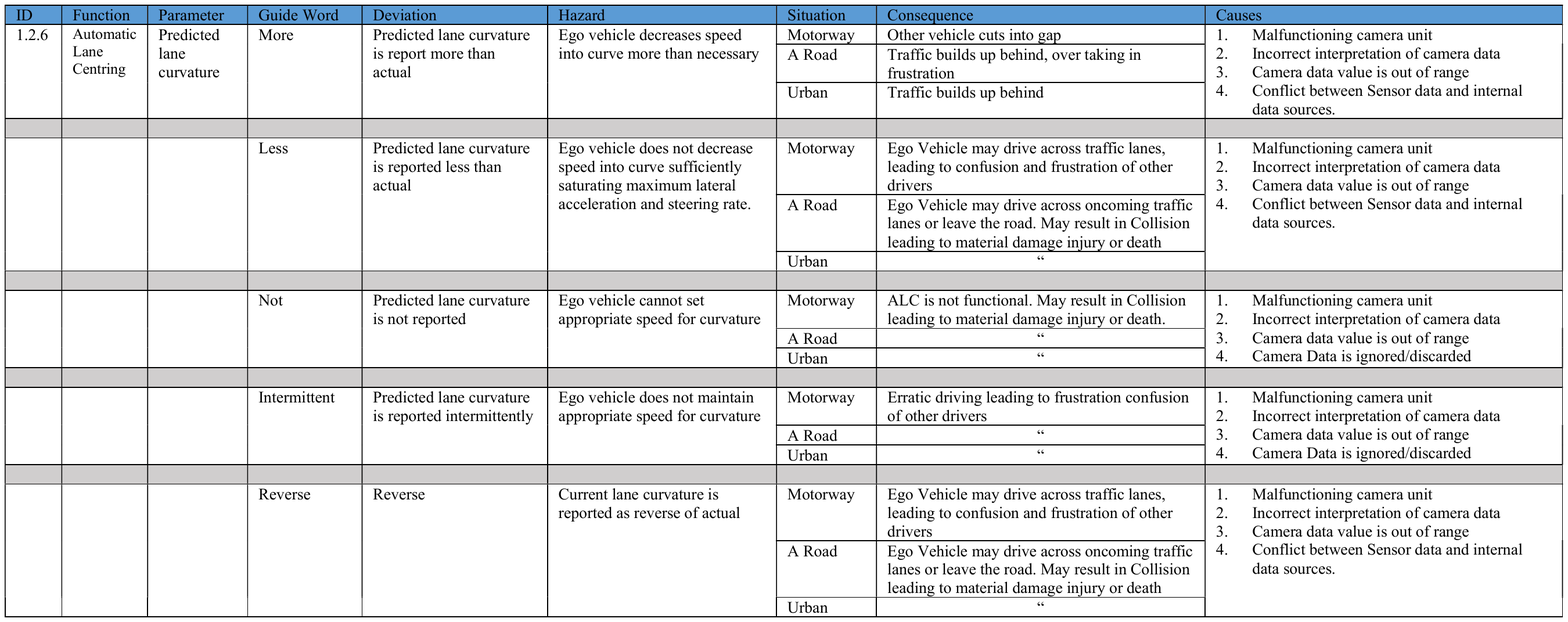}
  \caption{Hazop Table for camera implementation of \acrfull{alc} Part 5}
  \label{fig: Hazop_ALC_5}
\end{table}
\end{landscape}

\begin{table}[!h]
\caption{Scenario use-case: Radar implementation of Adaptive Cruise Control}
\begin{center}
\scalebox{0.70}{
\input{Tables/T_RDR_ACC_2_Table}}
\end{center}
\end{table}

\begin{landscape}
\begin{table}[ht]
  \centering
  \includegraphics[scale=0.75]{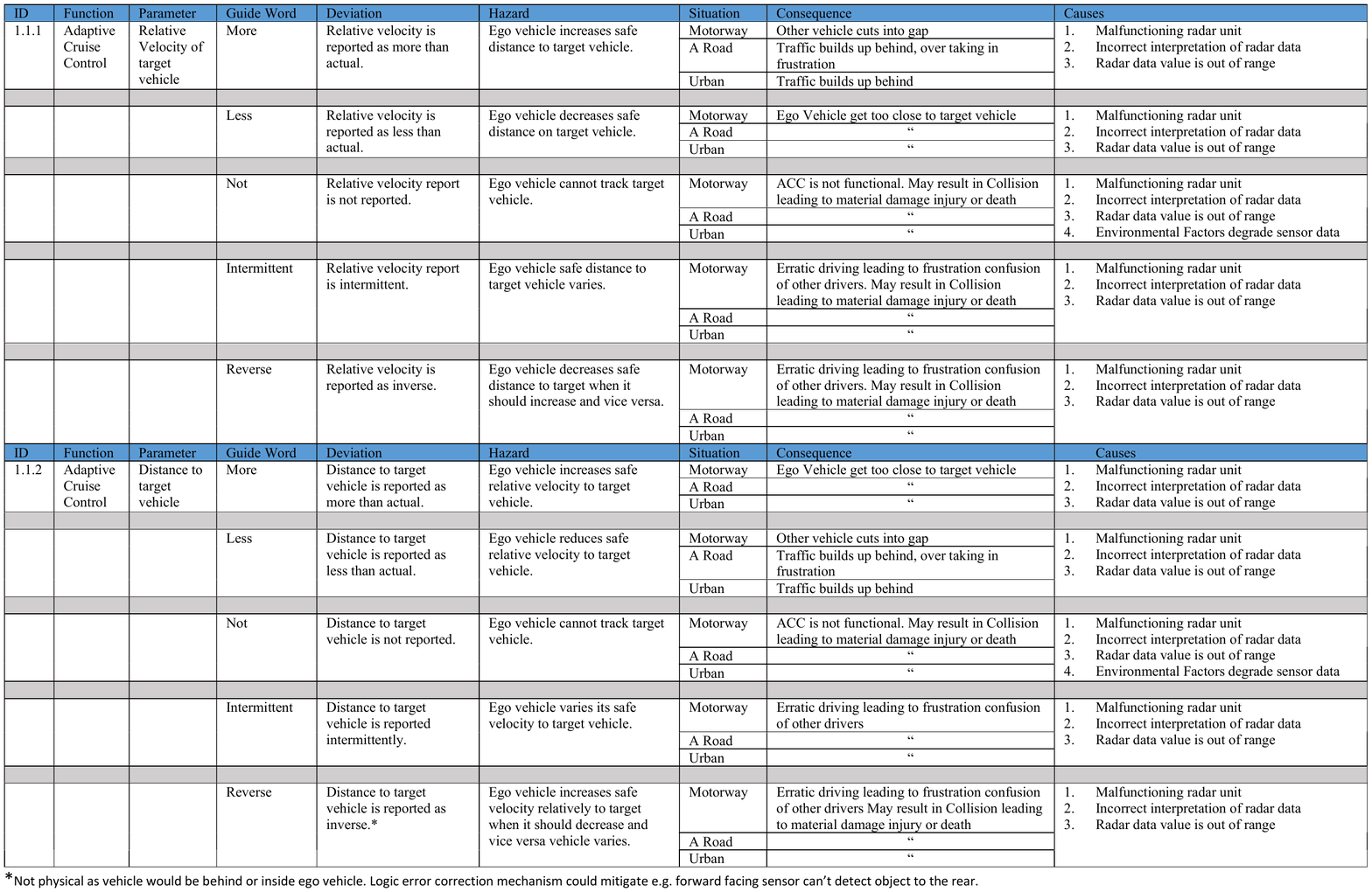}
  \caption{Hazop Table for radar implementation of \acrfull{acc} Part 1}
  \label{fig: Hazop_ACC_1}
\end{table}
\end{landscape}

\begin{landscape}
\begin{table}[ht]
  \centering
  \includegraphics[scale=0.75]{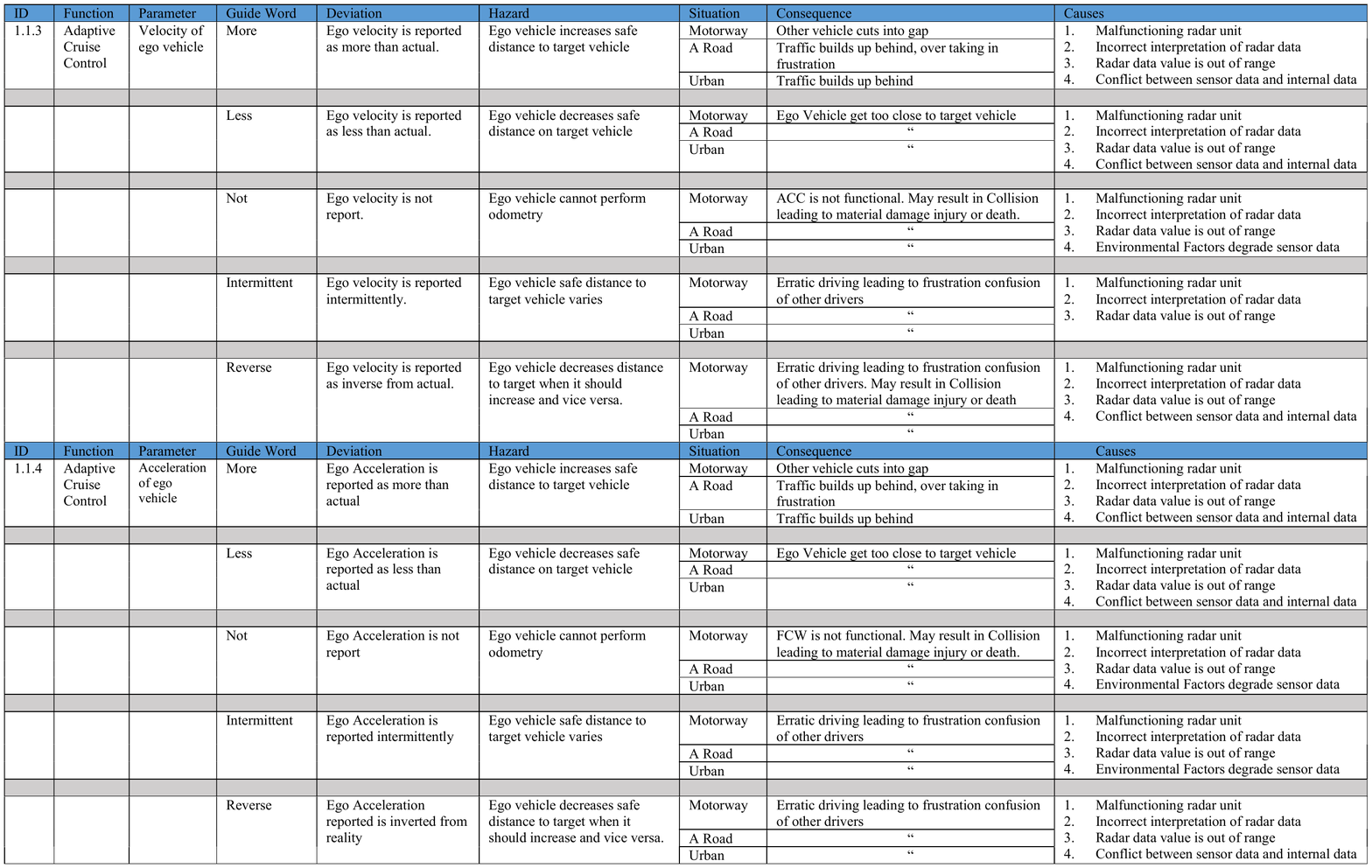}
  \caption{Hazop Table for radar implementation of \acrfull{acc} Part 2}
  \label{fig: Hazop_ACC_2}
\end{table}
\end{landscape}

%\begin{landscape}
%\begin{table}
%\resizebox{\columnwidth}{!}{
%\centering
%\input{Tables/HazopTable_T_RDR_ACC}}
%\end{table}
%\end{landscape}

\section{Discussion}
\label{sec: Discussion}

The method developed and illustrated in the paper is based on a well-understood \acrshort{hazop} method. The illustration of the method on \acrshort{acc} and \acrshort{alc} functionalities shows the feasibility of the approach, but also opens up some questions about its utility, scalability, etc.

First, it might be argued that the ``problems'' identified are obvious and would be ``caught anyway'' by competent designers or Tier 1 suppliers. Where this is true, the method serves to confirm the appropriateness of the component for its application. However, the examples of malfunctions of current \acrshort{av}s show that even if in principle these problems can be found, in practice they are making their way into deployed systems. 

Second, real-world deployments of \acrshort{av}s are likely to employ
multi-modal sensing systems. The analysis, as presented, focuses on single modalities. We see it as necessary to analyse the sensing modalities independently as they can be impaired and/or fail in different ways. What the method would allow us to do, if analysing a multi-modal sensing system, is to identify \acrshort{dsr}s which can be met by use of other sensing modes. This can be illustrated by considering illusions in more detail. 

The occurrence of pareidolia type phenomena are inherent in object recognition by \acrshort{cnn} and its extent and susceptibility will be related to the distribution of training data, \acrshort{nn} architecture, and available information in the scene  \cite{abbasFaceRecognitionFacial2019, nguyenDeepNeuralNetworks2015, nguyenMultifacetedFeatureVisualization2016, InceptionismGoingDeeper} Owing to its origin and frequency this phenomenon is considered within the label of misclassification.
However, susceptibility to geometric illusion e.g. Muller-Lyer type illusions, has been observed in several safety critical instances. The susceptibility of \acrshort{cnn} to these illusions and that they develop perception biases similar to humans, has been demonstrated by Ward et al (2019) \cite{wardExploringPerceptualIllusions2019}. We therefore suggest that the training data for \acrshort{as} should include significant contributions from less structured environments and specifically those with less rectilinear orthogonal forms in order to develop more nuanced cues for understanding of geometrical space. Typical geometric illusions may also be readily overcome by the use of high-fidelity active sensors as part of the perception suite. 

Third, the systems of interest are likely to have many operational modes and the effects of failures can alter between modes. For example, given an \acrshort{acc} function, several deviations, e.g. ``Less'' of ``relative velocity of target vehicle'' will be different in a ``drive off from standstill'' mode as opposed to a ``speed hold'' mode. The \acrshort{hazop} tables will need to be extended to include a column for operational modes. Further, \acrshort{dsr}s could include transitions between modes -- for example to a speed-limited variant of \acrshort{acc} due to the weather conditions impairing the \acrshort{av}'s sensors. 

Fourth is the issue of interoception. As well as classical fault monitoring and built-in test there is value in assessing sensor performance, in particular to determine whether or not the sensors are being impaired by the weather conditions, e.g. rain or snow in the case of optical sensors. This is a complex topic in itself but, for example, changes in statistics of the noise spectrum of the back-scattered light incident on an optical sensor, e.g. a Lidar, can indicate the presence of degrading environments e.g. rain or fog \cite{espineiraRealisticLiDARNoise2021, vriesmanExperimentalAnalysisRain2020, ritterDENSEEnvironmentPerception2019} and hence sensor impairment. We refer to this sort of interoception as sensor ``introspection''. Related \acrshort{dsr}s might be for change in \acrshort{av} mode, e.g. reduced velocity, and/or changes in priorities in sensing modality.

Fifth, there might be a concern about the scale or complexity of the example analysis, and what it means for analysis of industrial-scale systems, especially as it has been carried out a general functional level, not reflecting a specific implementation architecture. This is a concern with all such analyses, but one which we believe can be overcome, especially where capabilities are relatively ``standardised'' as seems likely to be the case with \acrshort{av}s. We explain this by comparison with what has happened with civil aircraft. The hazards for ``standardised'' capabilities are well-known (often recorded in hazard lists) but these are extended for innovative implementations. For example, many large commercial aircraft use landing gear with wheels that caster, while the Airbus A380's is actively steerable. This means that there are new hazards -- for example landing with the wheels not aligned with the direction of travel -- and this can be added to the hazard list. To do something analogous for \acrshort{av}s would require \acrshort{hazop}-like analysis for major capabilities which would be common across vehicles, which can be picked up and refined for a specific application. The level of analysis presented here, i.e. in terms of \acrshort{acc}, \acrshort{alc}, etc., may be ideal for producing such a transferable analysis, although it may be that working at the level of more primitive behavioural capabilities, e.g. lane change, negotiating a roundabout, may be more practical (these capabilities can be composed to make higher level services). This remains an issue for further study.  

Finally, it should be observed that, given the degree of interconnection between sensors, and various perception and understanding tasks, these systems are potentially liable to a significant number of common mode failures. Moreover, in deployed systems is has been observed that these failures have complex interactions \cite{aidrivrCanFSDBETA2020} that are not always commutative and associative. Further analysis of common mode failure in the perception and understanding capability is beyond the scope of this paper, but is an important area for further study.

\section{Conclusions}
\label{sec: Conclusions}

Many \acrshort{as} are heavily dependent on their sensing capabilities to provide safe and effective services. Whilst the importance of the perception systems has long been recognised there has not, to our knowledge, previously been a systematic approach to analysing the way in which the behaviour and failure of such perception systems could contribute to the safety of the \acrshort{as}. This paper is intended to be a first step towards filling that ``gap''.

Although our aim in proposing the method was to make generic, i.e. apply to a range of \acrshort{as}, it has been developed in the context of \acrshort{av}s. In particular the definition of \acrshort{hazop} guidewords is motivated by an analysis of reported failures of perception systems mainly, but not entirely, related to \acrshort{av}s. However, to show that the method is not specific to optical/camera-based sensing we have illustrated it on radar-based \acrshort{av} functionality.

To gain initial feedback on the use of the method we are applying it to a small-scale mobile robot, intended to provide simple delivery services in a University building. We are also seeking other opportunities to validate the method, as well as exploring some of the issues covered in the discussion such as sensor ``introspection'' and related issues such as sensor performance evaluation. Ultimately, our aim is to help provide a sounder basis for assessing the contribution of proposed sensing systems to safety, and to provide methods to help in the selection and refinement of sensing systems for \acrshort{as}.

\subsection*{Credit Authorship Statement}

All authors contributed equally to conception and drafting  of the manuscript

\subsection*{Declaration of Competing Interest}

The authors have no conflict of interest

\subsection*{Acknowledgment}

This work has been funded by Lloyds Register Foundation and the University of York through
the Assuring Autonomy International Programme https://www.york.ac.uk/assuring-autonomy.

\newpage
%see appendix~\ref{sec:sample:appendix}.

\appendix
\section{Supplementary HAZOP Tables}
\label{sec:appendix1}

\begin{table}[!ht]
\caption{Scenario use-case: Radar implementation of Automatic Emergency Braking}
\begin{center}
\scalebox{0.70}{
\input{Tables/T_RDR_AEB_3_Table}}
\end{center}
\end{table}

\begin{landscape}
\begin{table}[ht]
  \centering
  \includegraphics[scale=0.75]{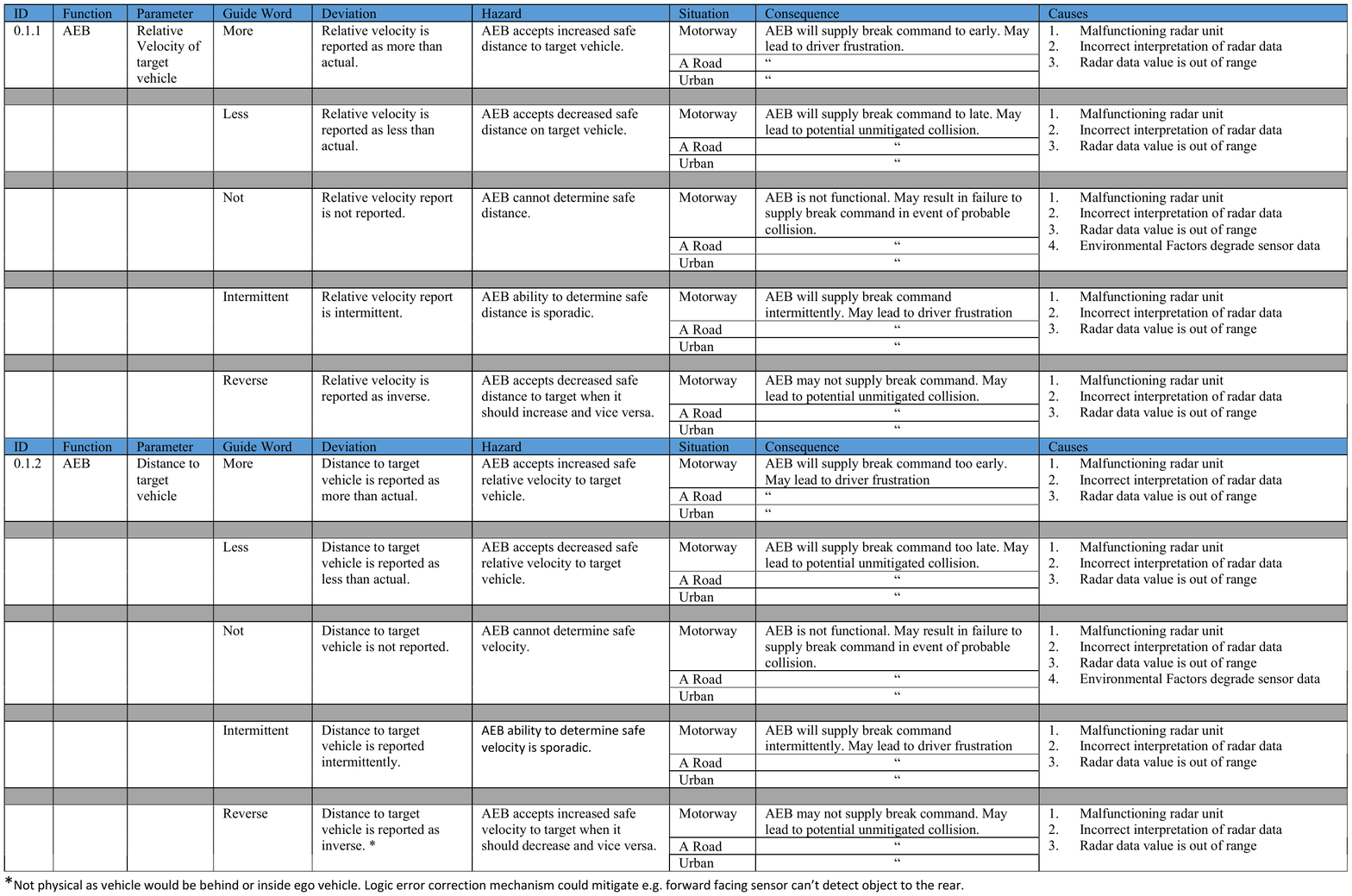}
  \caption{Hazop Table for radar implementation of \acrfull{aeb} Part 1}
  \label{fig: Hazop_AEB_1}
\end{table}
\end{landscape}

\begin{landscape}
\begin{table}[ht]
  \centering
  \includegraphics[scale=0.75]{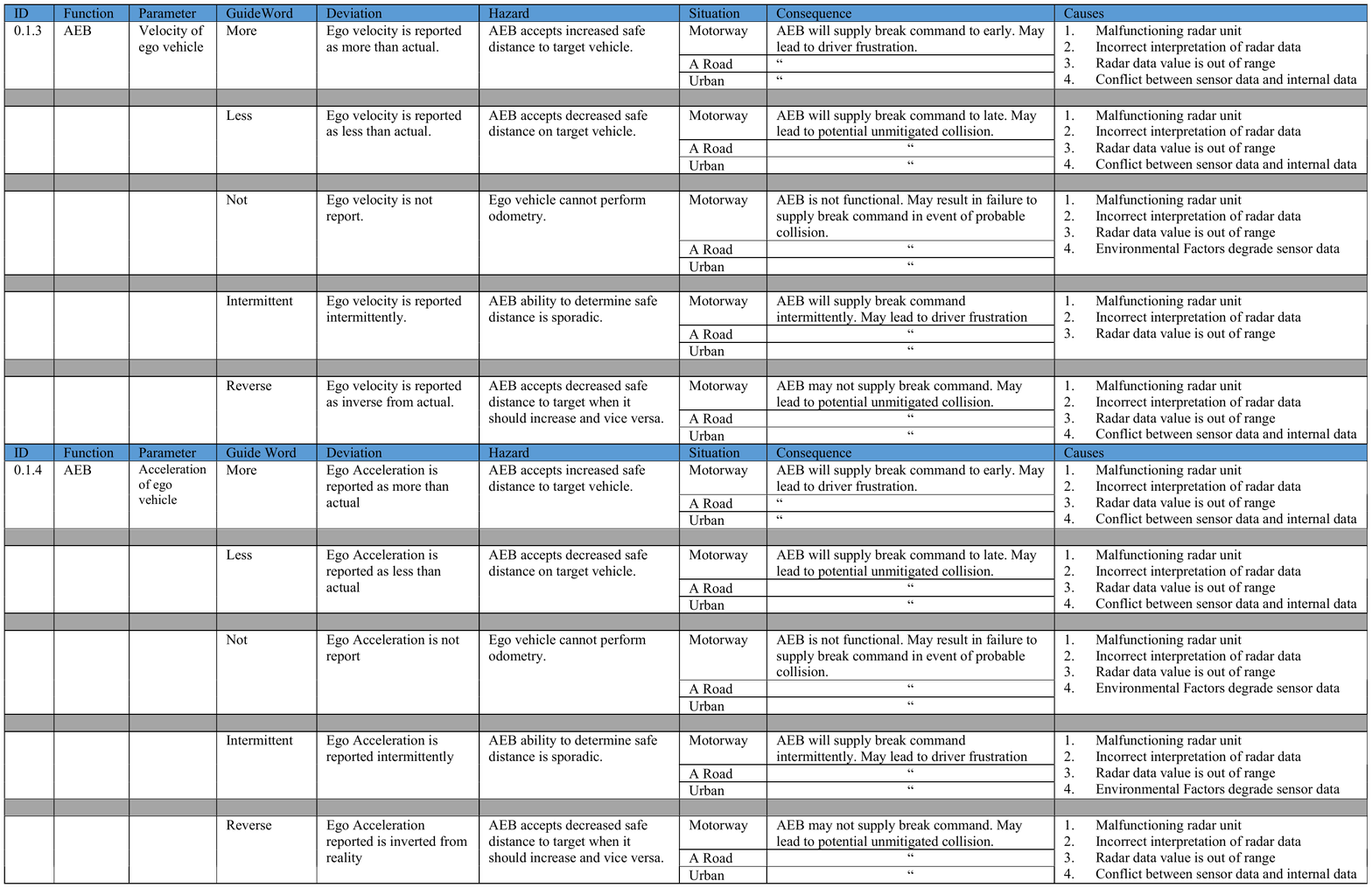}
  \caption{Hazop Table for radar implementation of \acrfull{aeb} Part 2}
  \label{fig: Hazop_AEB_2}
\end{table}
\end{landscape}

\begin{landscape}
\begin{table}[ht]
  \centering
  \includegraphics[scale=0.75]{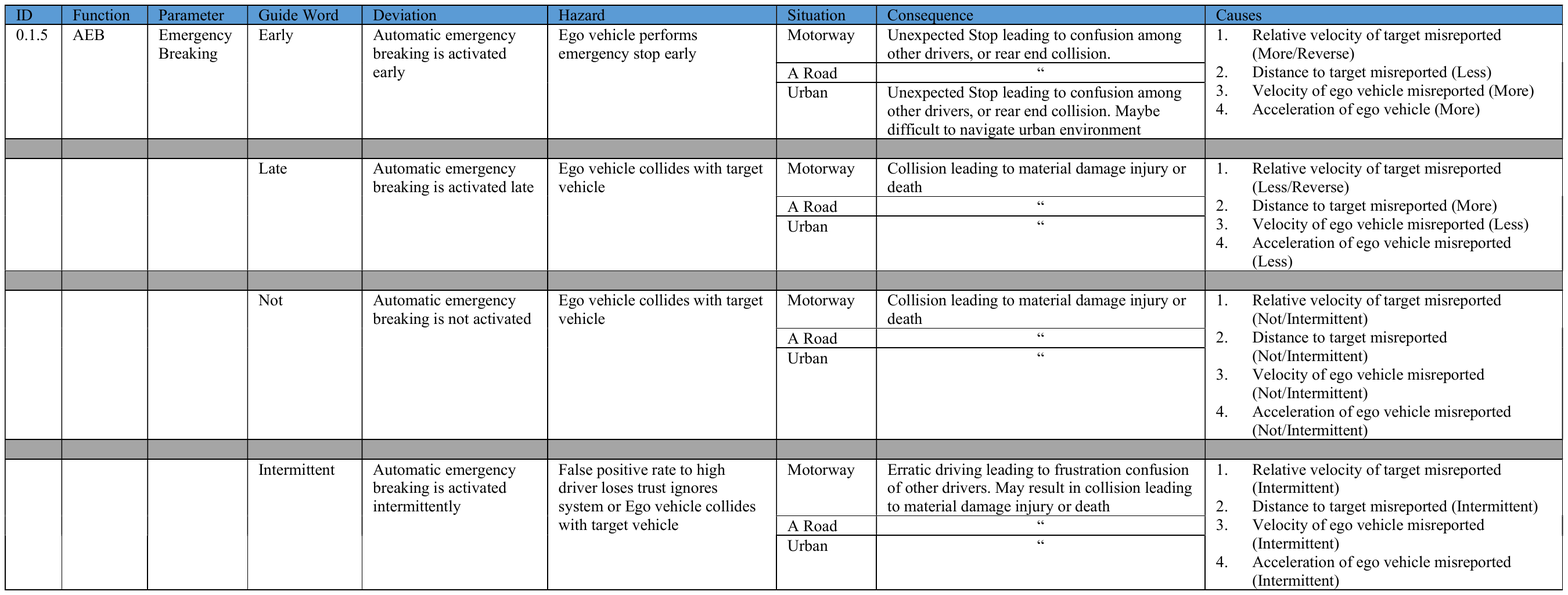}
  \caption{Hazop Table for radar implementation of \acrfull{aeb} Part 3}
  \label{fig: Hazop_AEB_3}
\end{table}
\end{landscape}

\begin{table}
\caption{Scenario use-case: Radar implementation of Forward Collision Warning}
\begin{center}
\scalebox{0.70}{
\input{Tables/T_RDR_FCW_4_Table}}
\end{center}
\end{table}

\begin{landscape}
\begin{table}[ht]
  \centering
  \includegraphics[scale=0.75]{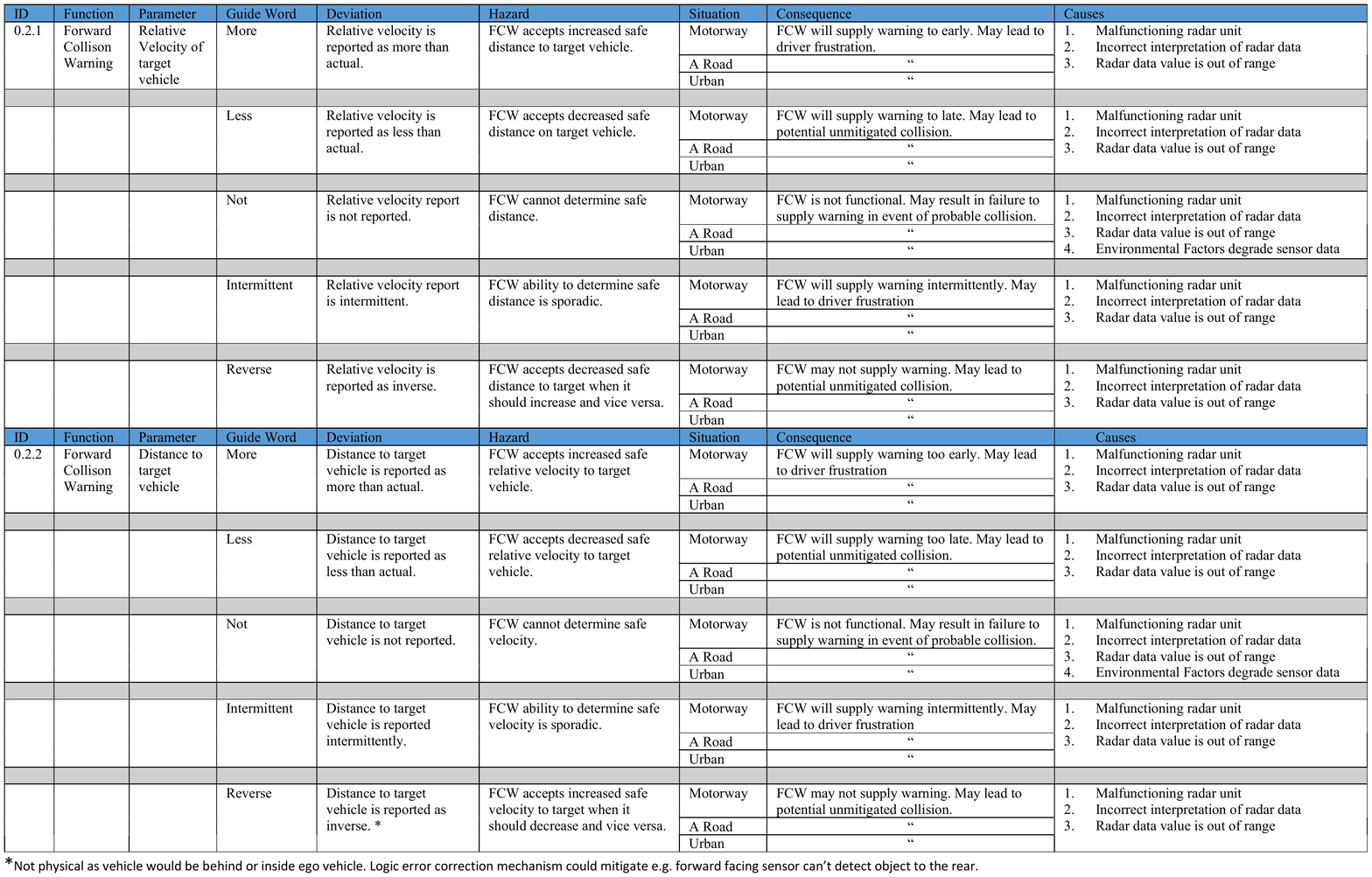}
  \caption{Hazop Table for radar implementation of \acrfull{fcw} Part 1}
  \label{fig: Hazop_FCW_1}
\end{table}
\end{landscape}

\begin{landscape}
\begin{table}[ht]
  \centering
  \includegraphics[scale=0.75]{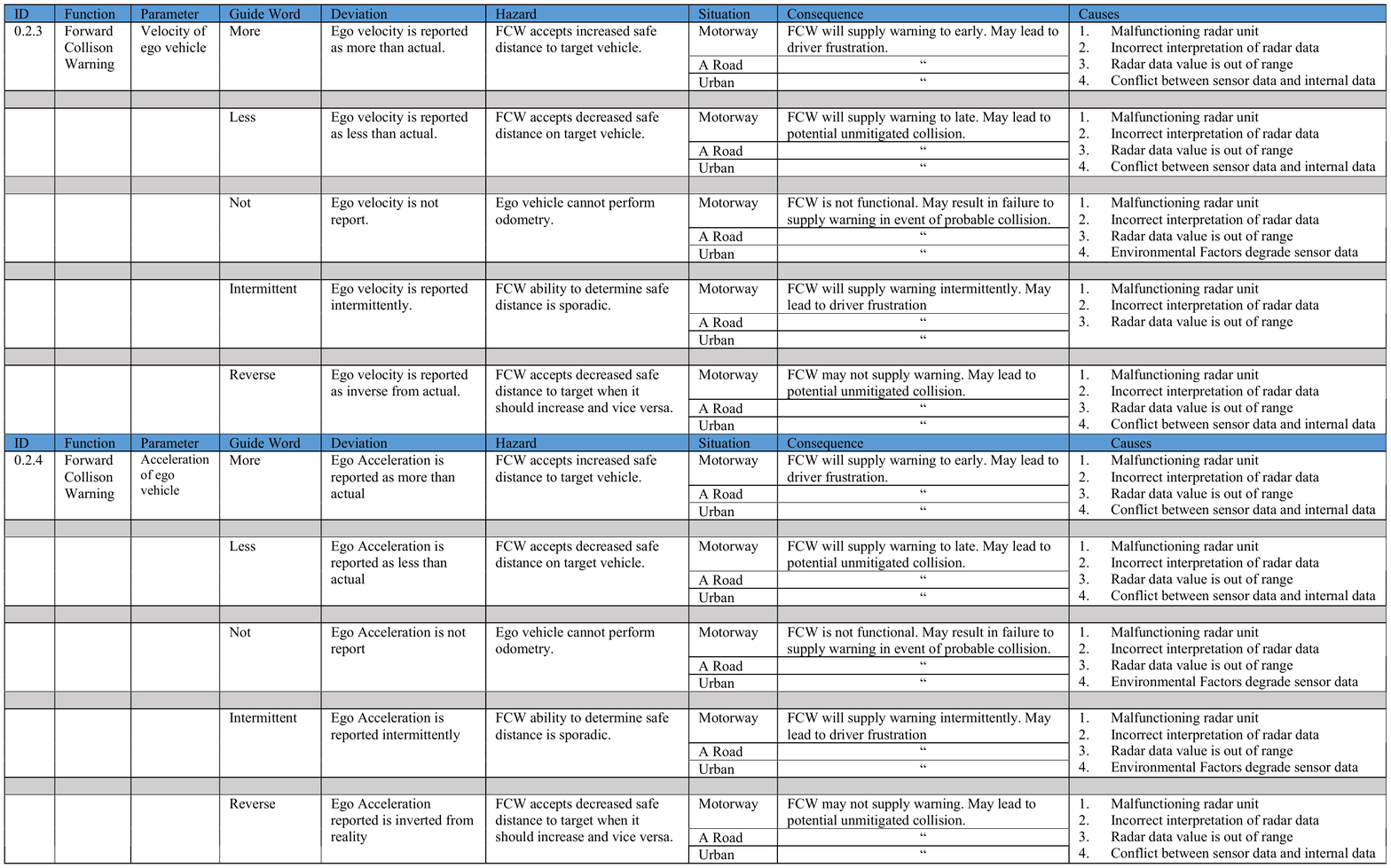}
  \caption{Hazop Table for radar implementation of \acrfull{fcw} Part 2}
  \label{fig: Hazop_FCW_2}
\end{table}
\end{landscape}

\begin{landscape}
\begin{table}[ht]
  \centering
  \includegraphics[scale=0.75]{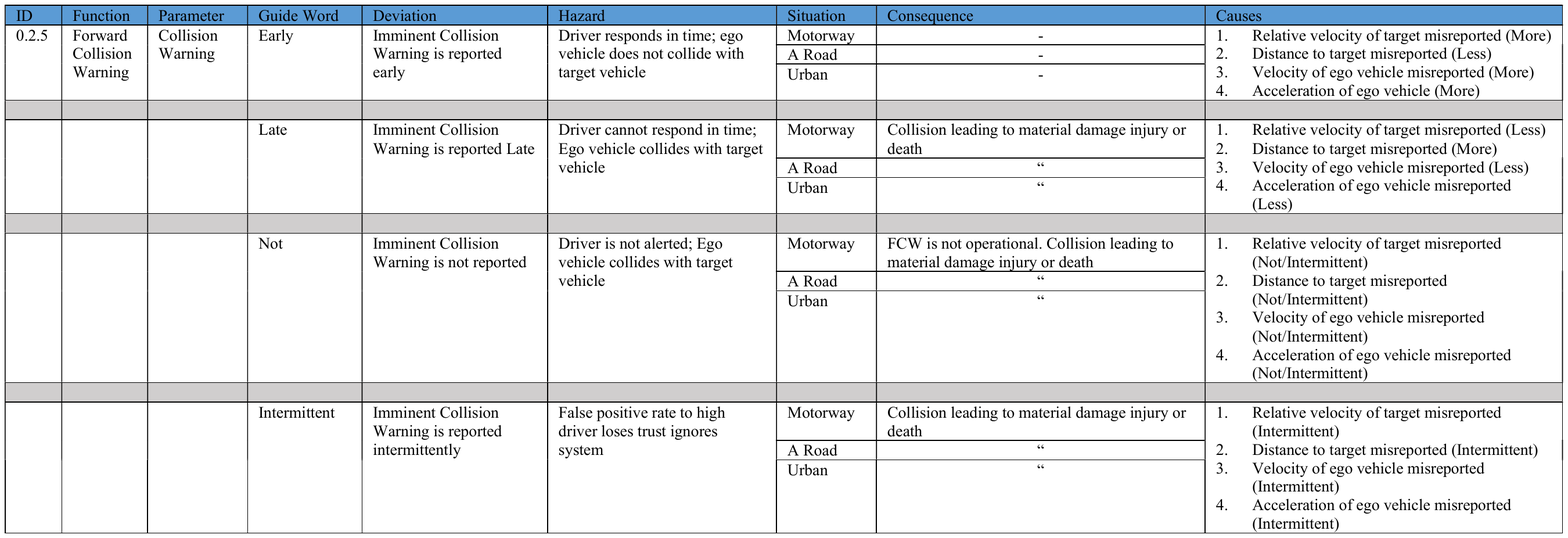}
  \caption{Hazop Table for radar implementation of \acrfull{fcw} Part 3}
  \label{fig: Hazop_FCW_3}
\end{table}
\end{landscape}

 \bibliographystyle{elsarticle-num} 
 \bibliography{SAUS}





% \end{thebibliography}

\clearpage
\printglossary[type=\acronymtype]
%%\printglossary

\end{document}

%% file: myacronyms.tex
\newacronym{acc}{ACC}{Adaptive Cruise Control}

\newacronym{acsf}{ACSF}{Automatically Commanded Steering Function}

\newacronym{ads}{ADS}{Automated Driving Systems}

\newacronym{adas}{ADAS}{Advanced Driver Assistance Systems}

\newacronym{aeb}{AEB}{Automatic Emergency Breaking}

\newacronym{alks}{ALKS}{Advanced Lane Keeping System}

\newacronym{alc}{ALC}{Automated Lane Centring}

\newacronym{amlas}{AMLAS}{Assurance of Machine Learning for Autonomous System}

\newacronym{as}{AS}{Autonomous Systems}

\newacronym{av}{AV}{Autonomous Vehicle}

\newacronym{bsw}{BSW}{Blind Sport Warning}

\newacronym{cnn}{CNN}{Convolutional Neural Network}

\newacronym{cots}{COTS}{commercial of the shelf}

\newacronym{ddm}{DDM}{Date Driven Model}

\newacronym{ddt}{DDT}{Dynamic Driving Task}

\newacronym{dsr}{DSR}{Derived Safety Requirment}

\newacronym{fcw}{FCW}{Forward Collision Warning}

\newacronym{fmea}{FMEA}{Failure Modes and Effects Analyses}

\newacronym{gpu}{GPU}{Graphics Processing Unit}

\newacronym{hazop}{HAZOP}{Hazard and Operability Studies}

\newacronym{hdr}{HDR}{High Dynamic Range}

\newacronym{hiss}{HISS}{Hazardous Internal System State}

\newacronym{ldw}{LDW}{Lane Departure Warning}

\newacronym{lka}{LKA}{Lane Keeping Assist}

\newacronym{ml}{ML}{Machine Learning}

\newacronym{nhtsa}{NHTSA}{ National Highway Traffic Safety Administration}

\newacronym{ntsb}{NTSB}{National Transportation Safety Board}

\newacronym{nn}{NN}{Neural Network}

\newacronym{odd}{ODD}{Operation Design Domain}

\newacronym{ota}{OTA}{Over-the-Air}

\newacronym{rgb}{RGB}{Red, Green, Blue}

\newacronym{sace}{SACE}{Safety Assurance of Autonomous Systems in Complex Environments}

\newacronym{sae}{SAE}{Society of Automotive 
Engineers}

\newacronym{saus}{SAUS}{Safety Assurance of Understanding in Autonomous Systems}

\newacronym{sc}{SC}{Safety Critical}

\newacronym{shard}{SHARD}{Software Hazard Analysis and Resolution in Design}

\newacronym{sotif}{SOTIF}{Safety  of  the  Intended  Function}

\newacronym{stpa}{STPA}{System-Theoretic Process Analysis}

\newacronym{unece}{UN ECE}{United Nations Economic Commission for Europe}

\newglossaryentry{latex}
{
        name=latex,
        description={Is a mark up language specially suited for 
scientific documents}
}

%% file: Tables/T_CMRA_ALC_1_Table.tex
% \usepackage{multirow}
\setlength{\arrayrulewidth}{0.5mm}
\setlength{\tabcolsep}{18pt}
\renewcommand{\arraystretch}{1.5}

\begin{tabular}{|llllllll|} \hline
\multicolumn{8}{|c|}{\cellcolor[HTML]{4C90CA} Hazop Analysis} \\ \hline
\multicolumn{3}{|l|}{\cellcolor[HTML]{4C90CA}Use-Case} & \multicolumn{5}{l|}{Camera Automatic Lane Centring} \\ \hline
\multicolumn{3}{|l|}{\cellcolor[HTML]{4C90CA}Use-Case \#} & \multicolumn{5}{l|}{T\_CMRA\_ALC\_1} \\ \hline
\multicolumn{8}{|l|}{\cellcolor[HTML]{4C90CA}Scenario} \\ \hline
\multicolumn{8}{|c|}{
    \begin{minipage}[c]{\textwidth}
    \centering
        \includegraphics[width=\columnwidth]{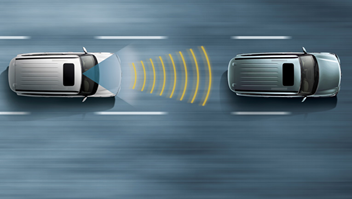} 
    \end{minipage}} \\\hline
\multicolumn{2}{|l|}{\cellcolor[HTML]{4C90CA}Primary environment} & \multicolumn{6}{l|}{Motorway, A   roads, urban} \\ \hline
\multicolumn{2}{|l|}{\cellcolor[HTML]{4C90CA}Goal in context} & \multicolumn{6}{l|}{System to detect and maintain vehicle in centre of the driving lane.} \\ \hline
\multicolumn{2}{|l|}{\cellcolor[HTML]{4C90CA}Scope} & \multicolumn{6}{l|}{} \\ \hline
\multicolumn{2}{|l|}{\cellcolor[HTML]{4C90CA}Pre-conditions} & \multicolumn{6}{l|}{Ego vehicle on a recognised lined carriage way}  \\ \hline
\multicolumn{2}{|l|}{\cellcolor[HTML]{4C90CA}Success end condition} & \multicolumn{6}{l|}{\parbox{12cm}{Ego vehicle adjusts velocity and steering angle consistent with safe maintenance of position in centre of driving lane}} \\ \hline
\multicolumn{2}{|l|}{\cellcolor[HTML]{4C90CA}Failed end conditions} & \multicolumn{6}{l|}{Ego vehicle leaves the driving lane and/or  carriage way} \\ \hline
\multicolumn{2}{|l|}{\cellcolor[HTML]{4C90CA}Actors} & \multicolumn{6}{l|}{Ego vehicle   control system} \\ \hline
\multicolumn{2}{|l|}{\cellcolor[HTML]{4C90CA}Trigger} & \multicolumn{6}{l|}{Ego vehicle navigating a recognised lined carriage way } \\ \hline
\multicolumn{2}{|l|}{\cellcolor[HTML]{4C90CA}Description} & \multicolumn{2}{l|}{\cellcolor[HTML]{4C90CA}Step} & \multicolumn{4}{l|}{\cellcolor[HTML]{4C90CA}Action} \\ \hline
\multicolumn{2}{|l|}{} & \multicolumn{2}{l|}{1} & \multicolumn{4}{l|}{Detect recognised lined carriage way} \\ \hline
\multicolumn{2}{|l|}{} & \multicolumn{2}{l|}{2} & \multicolumn{4}{l|}{\parbox{11cm}{Report velocity and steering angle command to vehicle control function}} \\ \hline
\multicolumn{2}{|l|}{\cellcolor[HTML]{4C90CA}Extension} & \multicolumn{2}{l|}{\cellcolor[HTML]{4C90CA}Step} & \multicolumn{4}{l|}{\cellcolor[HTML]{4C90CA}Branching Action} \\ \hline
\multicolumn{2}{|l|}{} & \multicolumn{2}{l|}{1}    & \multicolumn{4}{l|}{Failed to detect recognised lined carriage way} \\ \hline
\multicolumn{2}{|l|}{} & \multicolumn{2}{l|}{2}    & \multicolumn{4}{l|}{Report failure to vehicle operator} \\ \hline
\end{tabular}

%% file: Tables/T_RDR_ACC_2_Table.tex
% \usepackage{multirow}
\setlength{\arrayrulewidth}{0.5mm}
\setlength{\tabcolsep}{18pt}
\renewcommand{\arraystretch}{1.5}

\begin{tabular}{|llllllll|} \hline
\multicolumn{8}{|c|}{\cellcolor[HTML]{4C90CA} Hazop Analysis} \\ \hline
\multicolumn{3}{|l|}{\cellcolor[HTML]{4C90CA}Use-Case} & \multicolumn{5}{l|}{Radar Adaptive Cruise Control} \\ \hline
\multicolumn{3}{|l|}{\cellcolor[HTML]{4C90CA}Use-Case \#} & \multicolumn{5}{l|}{T\_RDR\_ACC\_2} \\ \hline
\multicolumn{8}{|l|}{\cellcolor[HTML]{4C90CA}Scenario} \\ \hline
\multicolumn{8}{|c|}{
    \begin{minipage}[c]{\textwidth}
    \centering
        \includegraphics[width=\columnwidth]{Figures/ForwardSensorMultiLane.png} 
    \end{minipage}} \\\hline
\multicolumn{2}{|l|}{\cellcolor[HTML]{4C90CA}Primary environment} & \multicolumn{6}{l|}{Motorway, A   roads, urban} \\ \hline
\multicolumn{2}{|l|}{\cellcolor[HTML]{4C90CA}Goal in context} & \multicolumn{6}{l|}{System to detect target and maintain safe following distance.} \\ \hline
\multicolumn{2}{|l|}{\cellcolor[HTML]{4C90CA}Scope} & \multicolumn{6}{l|}{} \\ \hline
\multicolumn{2}{|l|}{\cellcolor[HTML]{4C90CA}Pre-conditions} & \multicolumn{6}{l|}{Appearance of   Target in the road in front of ego vehicle}  \\ \hline
\multicolumn{2}{|l|}{\cellcolor[HTML]{4C90CA}Success end condition} & \multicolumn{6}{l|}{Ego vehicle adjusts velocity/distance consistent with target behaviour} \\ \hline
\multicolumn{2}{|l|}{\cellcolor[HTML]{4C90CA}Failed end conditions} & \multicolumn{6}{l|}{Ego vehicle   strikes target object or leaves the carriage way} \\ \hline
\multicolumn{2}{|l|}{\cellcolor[HTML]{4C90CA}Actors} & \multicolumn{6}{l|}{Ego vehicle   control system} \\ \hline
\multicolumn{2}{|l|}{\cellcolor[HTML]{4C90CA}Trigger} & \multicolumn{6}{l|}{Target in the   road that may not be driven over} \\ \hline
\multicolumn{2}{|l|}{\cellcolor[HTML]{4C90CA}Description} & \multicolumn{2}{l|}{\cellcolor[HTML]{4C90CA}Step} & \multicolumn{4}{l|}{\cellcolor[HTML]{4C90CA}Action} \\ \hline
\multicolumn{2}{|l|}{} & \multicolumn{2}{l|}{1} & \multicolumn{4}{l|}{Detect target on the road} \\ \hline
\multicolumn{2}{|l|}{} & \multicolumn{2}{l|}{2} & \multicolumn{4}{l|}{\parbox{11cm}{Report velocity and distance command to vehicle control function}} \\ \hline
\multicolumn{2}{|l|}{\cellcolor[HTML]{4C90CA}Extension} & \multicolumn{2}{l|}{\cellcolor[HTML]{4C90CA}Step} & \multicolumn{4}{l|}{\cellcolor[HTML]{4C90CA}Branching Action} \\ \hline
\multicolumn{2}{|l|}{} & \multicolumn{2}{l|}{1}    & \multicolumn{4}{l|}{Failure to adjust ego velocity or distance} \\ \hline
\multicolumn{2}{|l|}{} & \multicolumn{2}{l|}{2}    & \multicolumn{4}{l|}{Report failure to vehicle operator} \\ \hline
\end{tabular}

%% file: Tables/T_RDR_AEB_3_Table.tex
% \usepackage{multirow}
\setlength{\arrayrulewidth}{0.5mm}
\setlength{\tabcolsep}{18pt}
\renewcommand{\arraystretch}{1.5}

\begin{tabular}{|llllllll|} \hline
\multicolumn{8}{|c|}{\cellcolor[HTML]{4C90CA} Hazop Analysis} \\ \hline
\multicolumn{3}{|l|}{\cellcolor[HTML]{4C90CA}Use-Case} & \multicolumn{5}{l|}{Radar Automatic Emergency Break} \\ \hline
\multicolumn{3}{|l|}{\cellcolor[HTML]{4C90CA}Use-Case \#} & \multicolumn{5}{l|}{T\_RDR\_AEB\_3} \\ \hline
\multicolumn{8}{|l|}{\cellcolor[HTML]{4C90CA}Scenario} \\ \hline
\multicolumn{8}{|c|}{
    \begin{minipage}[c]{\textwidth}
    \centering
        \includegraphics[width=\columnwidth]{Figures/ForwardSensorMultiLane.png} 
    \end{minipage}} \\\hline
\multicolumn{2}{|l|}{\cellcolor[HTML]{4C90CA}Primary environment} & \multicolumn{6}{l|}{Motorway, A   roads, urban} \\ \hline
\multicolumn{2}{|l|}{\cellcolor[HTML]{4C90CA}Goal in context} & \multicolumn{6}{l|}{System to detect and avoid striking targets.} \\ \hline
\multicolumn{2}{|l|}{\cellcolor[HTML]{4C90CA}Scope} & \multicolumn{6}{l|}{} \\ \hline
\multicolumn{2}{|l|}{\cellcolor[HTML]{4C90CA}Pre-conditions} & \multicolumn{6}{l|}{Appearance of   Target in the road in front of ego vehicle}  \\ \hline
\multicolumn{2}{|l|}{\cellcolor[HTML]{4C90CA}Success end condition} & \multicolumn{6}{l|}{Ego vehicle slows or stops for target } \\ \hline
\multicolumn{2}{|l|}{\cellcolor[HTML]{4C90CA}Failed end conditions} & \multicolumn{6}{l|}{Ego vehicle   strikes target object or leaves the carriage way} \\ \hline
\multicolumn{2}{|l|}{\cellcolor[HTML]{4C90CA}Actors} & \multicolumn{6}{l|}{Ego vehicle   control system} \\ \hline
\multicolumn{2}{|l|}{\cellcolor[HTML]{4C90CA}Trigger} & \multicolumn{6}{l|}{Target in the   road that may not be driven over} \\ \hline
\multicolumn{2}{|l|}{\cellcolor[HTML]{4C90CA}Description} & \multicolumn{2}{l|}{\cellcolor[HTML]{4C90CA}Step} & \multicolumn{4}{l|}{\cellcolor[HTML]{4C90CA}Action} \\ \hline
\multicolumn{2}{|l|}{} & \multicolumn{2}{l|}{1} & \multicolumn{4}{l|}{Detect target on the road} \\ \hline
\multicolumn{2}{|l|}{} & \multicolumn{2}{l|}{2} & \multicolumn{4}{l|}{Report beak command to vehicle control function} \\ \hline
\multicolumn{2}{|l|}{\cellcolor[HTML]{4C90CA}Extension} & \multicolumn{2}{l|}{\cellcolor[HTML]{4C90CA}Step} & \multicolumn{4}{l|}{\cellcolor[HTML]{4C90CA}Branching Action} \\ \hline
\multicolumn{2}{|l|}{} & \multicolumn{2}{l|}{1}    & \multicolumn{4}{l|}{Detect target   on the road} \\ \hline
\multicolumn{2}{|l|}{} & \multicolumn{2}{l|}{2}    & \multicolumn{4}{l|}{Report target   to vehicle operator} \\ \hline
\end{tabular}

%% file: Tables/T_RDR_FCW_4_Table.tex
% \usepackage{multirow}
\setlength{\arrayrulewidth}{0.5mm}
\setlength{\tabcolsep}{18pt}
\renewcommand{\arraystretch}{1.5}

\begin{tabular}{|llllllll|} \hline
\multicolumn{8}{|c|}{\cellcolor[HTML]{4C90CA} Hazop Analysis} \\ \hline
\multicolumn{3}{|l|}{\cellcolor[HTML]{4C90CA}Use-Case} & \multicolumn{5}{l|}{Radar Forward Collision Warning} \\ \hline
\multicolumn{3}{|l|}{\cellcolor[HTML]{4C90CA}Use-Case \#} & \multicolumn{5}{l|}{T\_RDR\_FCW\_4} \\ \hline
\multicolumn{8}{|l|}{\cellcolor[HTML]{4C90CA}Scenario} \\ \hline
\multicolumn{8}{|c|}{
    \begin{minipage}[c]{\textwidth}
    \centering
        \includegraphics[width=\columnwidth]{Figures/ForwardSensorMultiLane.png} 
    \end{minipage}} \\\hline
\multicolumn{2}{|l|}{\cellcolor[HTML]{4C90CA}Primary environment} & \multicolumn{6}{l|}{Motorway, A   roads, urban} \\ \hline
\multicolumn{2}{|l|}{\cellcolor[HTML]{4C90CA}Goal in context} & \multicolumn{6}{l|}{System to detect and assist in avoiding striking a target.} \\ \hline
\multicolumn{2}{|l|}{\cellcolor[HTML]{4C90CA}Scope} & \multicolumn{6}{l|}{} \\ \hline
\multicolumn{2}{|l|}{\cellcolor[HTML]{4C90CA}Pre-conditions} & \multicolumn{6}{l|}{Appearance of   Target in the road in front of ego vehicle}  \\ \hline
\multicolumn{2}{|l|}{\cellcolor[HTML]{4C90CA}Success end condition} & \multicolumn{6}{l|}{Ego vehicle slows or stops for target.} \\ \hline
\multicolumn{2}{|l|}{\cellcolor[HTML]{4C90CA}Failed end conditions} & \multicolumn{6}{l|}{Ego vehicle   strikes target object or leaves the carriage way} \\ \hline
\multicolumn{2}{|l|}{\cellcolor[HTML]{4C90CA}Actors} & \multicolumn{6}{l|}{Ego vehicle   control system} \\ \hline
\multicolumn{2}{|l|}{\cellcolor[HTML]{4C90CA}Trigger} & \multicolumn{6}{l|}{Target in the   road that may not be driven over} \\ \hline
\multicolumn{2}{|l|}{\cellcolor[HTML]{4C90CA}Description} & \multicolumn{2}{l|}{\cellcolor[HTML]{4C90CA}Step} & \multicolumn{4}{l|}{\cellcolor[HTML]{4C90CA}Action} \\ \hline
\multicolumn{2}{|l|}{} & \multicolumn{2}{l|}{1} & \multicolumn{4}{l|}{Detect target on the road} \\ \hline
\multicolumn{2}{|l|}{} & \multicolumn{2}{l|}{2} & \multicolumn{4}{l|}{Report target   to vehicle operator} \\ \hline
\multicolumn{2}{|l|}{\cellcolor[HTML]{4C90CA}Extension} & \multicolumn{2}{l|}{\cellcolor[HTML]{4C90CA}Step} & \multicolumn{4}{l|}{\cellcolor[HTML]{4C90CA}Branching Action} \\ \hline
\multicolumn{2}{|l|}{} & \multicolumn{2}{l|}{1}    & \multicolumn{4}{l|}{Vehicle operator does not initiate breaking} \\ \hline
\multicolumn{2}{|l|}{} & \multicolumn{2}{l|}{2}    & \multicolumn{4}{l|}{Report warning command to vehicle control function (Trigger AEB)} \\ \hline
\end{tabular}